\documentclass[aps, prd, onecolumn, superscriptaddress, nofootinbib, amsmath, amssymb]{revtex4-2}
\usepackage{graphicx}
\usepackage{amssymb}
\usepackage{amsmath}
\usepackage{bm}
\usepackage{xcolor}

\newcommand{\rf}[1]{(\ref{#1})}
\newcommand{\beq}{\begin{equation}}
\newcommand{\beql}[1]{\beq\label{#1}}
\newcommand{\eeq}{\end{equation}}
\newcommand{\bea}{\begin{eqnarray}}
\newcommand{\eea}{\end{eqnarray}}

\newcommand{\e}{\mbox{e}}

\newcommand{\G}{\Gamma}
\newcommand{\lam}{\lambda}
\newcommand{\Lam}{\Lambda}

\renewcommand{\a}{\alpha}

\newcommand{\om}{\omega}

\newcommand{\del}{\delta}
\newcommand{\Del}{\Delta}

\newcommand{\kp}{\kappa}
\newcommand{\oh}{\frac{1}{2}}

\newcommand{\ra}{\rangle}
\newcommand{\la}{\langle}

\newcommand{\prt}{\partial}
\newcommand{\mi}{\!-\!}
\newcommand{\equ}{\!=\!}
\newcommand{\plu}{\!+\!}

\newcommand{\cD}{{\cal D}}

\newcommand{\const}{\mathrm{const.}}
\newcommand{\UV}{\mathrm{\scriptscriptstyle UV}}
\newcommand{\IR}{\mathrm{\scriptscriptstyle IR}}

\usepackage{ulem}                   
\begin{document}

\title{Is Lattice Quantum Gravity  Asymptotically Safe ? \\
Making contact between Causal Dynamical Triangulations\\ and the Functional Renormalization Group.}

\author{J. Ambj\o rn}
\email{ambjorn@nbi.dk,ambjorn@science.ru.nl}
\affiliation{The Niels Bohr Institute, Copenhagen University Blegdamsvej 17, DK-2100 Copenhagen \O , Denmark.}
\affiliation{Institute for Mathematics, Astrophysics and Particle Physics (IMAPP) Radboud University Nijmegen, Heyendaalseweg 135, 6525 AJ  Nijmegen, The Netherlands}

\author{J. Gizbert-Studnicki}
\email{jakub.gizbert-studnicki@uj.edu.pl}
\author{A. G\"{o}rlich}
\email{andrzej.goerlich@uj.edu.pl}
 \affiliation{Institute of Theoretical Physics, Jagiellonian University, ul. prof. S. \L ojasiewicza 11, Krak\'ow, PL 30-348, Poland.}
\affiliation{Mark Kac Center for Complex Systems Research, Jagiellonian University, ul. prof. S. \L ojasiewicza 11, Krak\'ow, PL 30-348, Poland.}

\author{D. Németh}
\email{nemeth.daniel.1992@gmail.com}
\affiliation{Institute for Mathematics, Astrophysics and Particle Physics
(IMAPP) Radboud University Nijmegen, Heyendaalseweg 135, 6525 AJ  Nijmegen, The Netherlands}

\begin{abstract}
We compare the effective action of the scale factor obtained from lattice quantum gravity (in the form of Causal Dynamical Triangulations (CDT)) to the corresponding effective action obtained from the simplest Functional Renormalization Group (FRG) calculation. In this way, we can identify the generic infinite four-volume limit of the lattice theory in the so-called de Sitter phase with the Gaussian fixed-point limit or the IR fixed-point limit obtained by FRG. We also show how to identify a putative UV lattice gravity fixed point. Our Monte Carlo simulations of CDT allow for the existence of such a UV fixed point, although the data precision does not yet provide a proof of its existence. The concept of a correlation length relevant for the lattice gravity fixed points is argued to be different from the concept of correlation lengths encountered in field theories in a fixed spacetime background. 
\end{abstract}

\maketitle

\newpage

\section{Introduction}
\label{intro} Four-dimensional Causal Dynamical Triangulations (CDT)  attempts to provide a lattice regularization of four-dimensional quantum gravity (see \cite{review1,review2} for reviews). The path integral is defined as a sum over a certain class of piecewise linear geometries and the length $a$ of the links of the triangulations defining the piecewise linear geometries acts as a UV cut-off, precisely as for ordinary lattice field theories. The main difference between ordinary lattice field theories and CDT is that the lattice in the former case is fixed, the dynamics coming from the fields that ``live'' on the lattice, while for CDT the dynamics originates from summing over different lattices, representing different geometries in the path integral.

The CDT program assumes that there exists a continuum 4d quantum field theory of gravity, and the task is then to study the lattice theory and show that one can fine tune the dimensionless bare coupling constants of the lattice theory in such a way that one can take the lattice cut-off $a \to 0$ and still obtain an interacting quantum theory.  This is a non-trivial task even for a standard renormalizable quantum field theory, and here it becomes even more difficult since quantum gravity is not a perturbatively renormalizable quantum field theory in four dimensions. The possibility that there might exist such a non-perturbatively renormalizable theory of gravity is known as the asymptotic safety scenario \cite{weinberg}. A lot of work has gone into testing this scenario using what is known as the Functional Renormalization Group approach (FRG), and there is good evidence that such a scenario could be true (see \cite{review3,review4,review5} for extensive reviews). However, even if there exists a non-perturbative UV fixed point where one can define a theory of quantum gravity, it is not clear that the corresponding theory will be a unitary theory. The unitarity problem is even older than the asymptotic safety scenario, since one can add $R^2$ terms to the Einstein-Hilbert action and then the theory can be formulated as a renormalizable theory \cite{stelle}, but the problem is that it will in general be a non-unitary theory and this is one reason it was originally not considered as a viable quantum gravity candidate. 

When solving the FRG equations in order to locate the non-perturbative UV fixed point one has to truncate the equations and it is difficult to judge how reliable such truncations are. It is therefore very desirable to verify the existence of the UV fixed point by an independent approach, and lattice gravity in the form of CDT is such an approach. It is based on Monte Carlo simulations and thus it will also involve approximations, but of a quite different nature than those used in the FRG equations. Also, it should be mentioned that the lattice CDT theory is a unitary theory\footnote{More precisely, after rotating to Euclidean signature in CDT, one can define a transfer matrix with respect to the CDT Euclidean time and show that it is reflection positive \cite{ajl2001}. In lattice field theories this ensures that one will obtain a unitary time evolution when rotating back to the Lorentzian signature.}. Thus, if the lattice results can confirm the FRG picture, it provides evidence in favor of the non-perturbative theory defined at the (putative) UV fixed point being a unitary theory. Also, for the same reason it is unlikely that the continuum limit of the lattice theory, should it be non-trivial, can be viewed as a continuum quantum gravity theory with a simple $R^2$ term added to the Einstein-Hilbert terms.

The lattice action used in the CDT lattice theory depends on three dimensionless coupling constants.
In this coupling constant space there are critical surfaces, defining phase transitions of the lattice theory. These critical surfaces intersect at critical lines and a UV fixed point could be positioned on such critical lines. The purpose of this article is to report on Monte Carlo measurements close to some of the critical lines and to compare the effective action obtained from the Monte Carlo simulations with the effective action obtained by continuum FRG calculations.

\section{FRG}

The very simplest effective action obtained by the FRG is a truncation where one considers the Einstein-Hilbert action with the gravitational coupling constant $G$ and the cosmological constant  $\Lam$ as functions of the so-called  scale $k$ entering the FRG equations. The corresponding effective action in spacetimes with Euclidean signature can then be written as 
\beq\label{j1}
\G_k[g_{\mu\nu}] = \frac{1}{16\pi G_k} \int d^4 x \sqrt{g(x)} \, \Big( -R(x) + 2 \Lam_k\Big)
\eeq
The running coupling constants are conjectured to have an UV fixed point for $k \to \infty$,
where they behave as
\beq\label{j2}
G_k := g_k/k^2, \quad g_k \to g_*, \qquad \Lam_k := \lam_k k^2, \quad \lam_k \to \lam_*,
\eeq  
where $g_k$ and $\lam_k$ are dimensionless coupling constants that one might try to compare to suitable dimensionless lattice gravity coupling constants. In particular we have for the dimensionless combination $G_k \Lam_k$:
\beq\label{j3}
G_k \Lam_k \to g_*\lam_*\quad {\rm for} \quad k \to \infty.
\eeq
Let us treat \rf{j1} as a standard effective action\footnote{In the actual FRG calculations one is often making the decomposition $g_{\mu\nu} = g_{\mu\nu}^B + h_{\mu\nu}$, where $g_{\mu\nu}^B$ is a fixed background metric (e.g. a fixed de Sitter metric) that is fixed even when the scale $k$ is changing. From first principles the effective action can only depend in $g_{\mu\nu}$, not the arbitrary choice $g_{\mu\nu}^B$ Our treatment here is the most naive implementation of what is suggested in \cite{fr1,fr2}, namely that the background one should use for a given scale $k$ should be the one that satisfy the equations of motion at that scale. In  \cite{fr2} it is called the choice of self-consistent background geometries.}. Then an extremum for  $\G_k[g_{\mu \nu}]$ is a de Sitter universe with cosmological constant $\Lam_k$. We will assume the metric is Euclidean and then the solution is a four-sphere, $S^4$, with radius $R_k = 3/\sqrt{\Lam_k}$. One can now study fluctuations around this solution. We will only do that here in the simplest possible way where we use a minisuperspace version of \rf{j1}. The reason for this restriction is that we want to compare with computer simulations of quantum gravity, in the version of CDT, where such a minisuperspace effective action appears when one integrates (numerically, via Monte Carlo simulations) over all degrees of freedom except the scale factor
$r(t)$ or the three-volume $V_3(t)$
defined in eqs.\ \rf{j5} and \rf{j6a}\footnote{In \cite{knorr-frank} it is shown that when calculating fluctuations for ``global'' quantities like the three-volume, only constant modes contribute when space is compact. These modes are precisely the  modes used when calculating fluctuations in the minisuperspace approximation.}.  Close to the UV fixed point we then have 
\beq\label{j4}
R_k = \frac{3}{\sqrt{\lam_k}}\, \frac{1}{k} \to \frac{3}{\sqrt{\lam_*}}\, \frac{1}{k}, 
\qquad V_4(k) = \frac{8 \pi^2}{3} R_k^4 = \frac{8 \pi^2}{3} \frac{81}{\lam_k^2} \frac{1}{k^4}
\to \frac{8 \pi^2}{3} \frac{81}{\lam_*^2} \frac{1}{k^4},
\eeq
i.e. the volume $V_4(k)$ of the de Sitter sphere goes to zero when approaching the UV fixed point. Does it make sense to study fluctuations ``around''  such small universe? Somewhat surprisingly the answer is affirmative because the coupling constant appearing in the study of fluctuations is $\sqrt{G_k \Lam_k} = \sqrt{g_k \lam_k}$ that is never large, even at the UV fixed point where $R_k$ formally is zero. In order to compare with computer simulations let us consider fluctuations around a de Sitter sphere with fixed volume $V_4$ rather than fixed $\Lam$. The minisuperspace action written  using the metric
\beq\label{j5}
ds^2 = dt^2 + r^2(t) d \Omega_3^2, 
\eeq
where $t$ denotes a choice of proper time and $d\Omega^2_3$ the 
squared line element on the unit three-sphere, is
\beq\label{j6}
S = -\frac{1}{24 \pi G_k} \int dt \,\Big( \frac{\dot{V}_3^2}{V_3} + \del_0 V_3^{1/3} \Big),
\qquad \int dt V_3(t) = V_4(k),
\eeq
where 
\beq\label{j6a}
V_3(t) = 2 \pi^2 r^3(t), \quad \del_0 = 9 (2\pi^2)^{2/3}.
\eeq
Introducing dimensionless variables $v_3 = V_3/V_4^{3/4}$
and $s = t/V_4^{1/4}$ we can write 
\beq\label{j8} 
S = -\frac{1}{24\pi} \frac{\sqrt{V_4(k)}}{G_k} 
\int ds\;\Big(  \frac{ \dot{v}_3^2}{v_3} + \del_0 v_3^{1/3} \Big), \qquad \int ds\, v_3(s) = 1.
\eeq
Here $s$ and $v_3(s)$ will be of order $O(1)$ and the classical solution (the 
four-sphere with volume 1) is 
\beq\label{j8a}
v_3^{cl}(s) = \frac{3}{4 \om_0} \cos^3 \Big(\frac{s}{\om_0} \Big), 
\quad \om_0^4 = \frac{3^4}{2^2} \frac{1}{\del_0^{3/2}} =\frac{3}{8\pi^2},
\quad s \in \Big[- \frac{\pi \om_0}{2},  \frac{\pi \om_0}{2}\Big].
\eeq
The fluctuations around $v_3^{cl}(s)$ will for a given $k$ be governed by the effective coupling constant
\beq\label{j9}
g_{\rm eff}^2(k) =\frac{ 24 \pi \,G_k}{ \sqrt{V_4(k)}} = \frac{4}{\sqrt{6}} \,\Lam_k G_k
\simeq 1.63 \, \lam_k g_k.
\eeq
In the FRG analysis $\lam_k g_k$ will be running from the present days value for small $k = k_p$ ($\lam_{k_p}g_{k_p} \approx 10^{-120}$) to the value $\lam_* g_*$. This UV fixed point value is not really universal,
but with a number of different regularizations one finds values like $\lam_* g_* = 0.12 \pm 0.02$ (see \cite{review3}, Table 2);
on the other hand, a recent calculation, including wave function renormalization in the calculation and using a different regularization,
finds $\lam_*g_* = 0.74 \, 10^{-3}$ \cite{kawai}, i.e. a value quite a lot lower\footnote{An earlier calculation \cite{kevin}, essentially using the same framework as in \cite{kawai}, but more general, obtained more or less agreement with the old FRG calculations, using the ``standard'' FRG regularization.}.
In both cases there is no obvious obstruction to use lowest order perturbative expansion around $v_3^{cl}$.
Denoting the fluctuations around $v_3^{cl}(s)$ by $x(s)$ we have (see \cite{agjl1} and 
\cite{review1} for details) 
\beq\label{j10}
v_3(s) = v_3^{cl}(s) + x(s) , \quad \frac{x(s)}{v_3^{cl}(s)} = O(g_{\rm eff}(k)).
\eeq
If we choose the two values of $\lam_* g_*$ mentioned above  we obtain
\beq\label{j11}
\frac{x(0)}{v_3^{cl}(0)} = {O(g_{\rm eff}(k))} \quad {\rm where}\quad
{g_{\rm eff}(k)} < 0.44 \textrm{ or } 0.035
\eeq 
for all values of $k$, starting at the present day small value $k_p$, to $k \to \infty$ when
approaching the UV fixed point.
It is remarkable that such a ``semiclassical'' description seems to be 
valid all the way to the UV fixed point. 

\section{The CDT Monte Carlo simulations}

\begin{figure}[t]
\centerline{\scalebox{1.0}{\rotatebox{0}%{\includegraphics{Phase_structure.pdf}}}} 
{\includegraphics{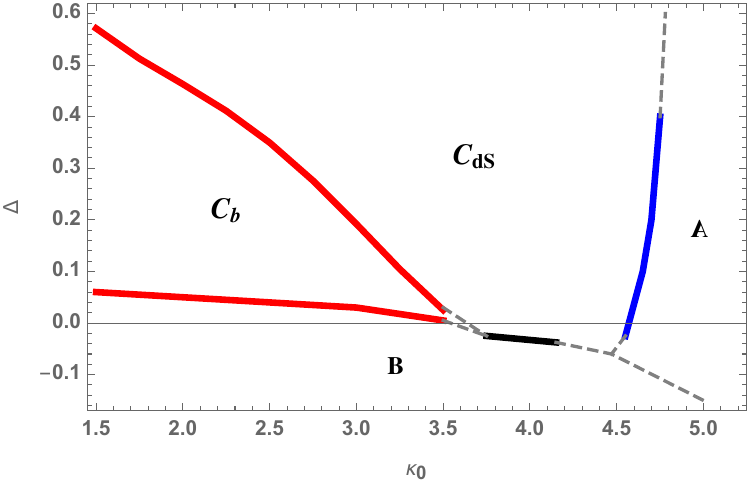}}}} 
\caption{The CDT phase diagram. Phase transition between phases $C_{\rm dS}$ and $C_b$ is (most likely) second order, as is the transition between $C_b$ and $B$, while the transition between $C_{\rm dS}$ and $A$ is first order. The transition between $C_{\rm dS}$ and $B$ is still under investigation.}
\label{fig1}
\end{figure}

The piecewise linear geometries (the lattice geometries) we will consider below are constructed from building blocks (four-simplices) glued together in various ways The length of the links, $a$ act as a UV cut off in the same way as for ordinary lattice field theories. Before starting a detailed discussion of the class of geometries we consider, let us just assume we manage to create geometries on the computer that are representatives of the geometries close to the FRG UV fixed point. Let us further assume that close to the fixed point we loosely can identify $a \propto 1/k$. A piecewise linear lattice geometry will then be constructed from $N_4$ four-simplices, and each of these will have a four-volume $\sqrt{5}/96 \cdot a^4$, i.e. a total four-volume of $V_4 = \sqrt{5}/96 \cdot N_4a^4 $. We can now compare this to the FRG expression \rf{j4} for $V_4(k)$ and we obtain
\beq\label{jw1}
N_4 \propto \frac{96}{\sqrt{5}}\, \frac{8 \pi^2}{3} \,
\frac{81}{\lam_*^2} 
\quad (\approx 2 \cdot  10^6).
\eeq
The approximate value is obtained by using a typical value for $\lam_*$ obtained in the FRG studies. The important point before starting the detailed discussion below is that if we write $a \propto 1/k$ close to the UV fixed point, we are led to a finite $N_4$ at the fixed point and we need non-trivial scaling in the lattice theory to avoid this conclusion. Below, more elaborate arguments lead to the same conclusion even if we do not assume $a \propto 1/k$ near the UV fixed point. The geometries one observes in the computer simulations depend on the choice of the bare coupling constants used in the lattice action. We denote these $\kp_0$, $\Del$, and $\kp_4$. We refer to \cite{review1} for a detailed discussion of the origin and interpretation of these, but have for completeness included a little summary of the notation in Appendix 1. Here we only remark that $\kp_0$ is related to the inverse bare dimensionless gravitational coupling constant, i.e.\ we have 
\beq\label{p1}
\kp_0 \propto \frac{a^2}{G},
\eeq
where $G$ denotes the gravitation coupling constant that 
appears in the Einstein-Hilbert action, while $\kp_4$ is related to the dimensionless cosmological coupling constant $a^4 \Lam/G$. Finally $\Del$ can be related to a possible asymmetry between space and time inherent in the CDT lattice implementation of spacetime. For given values $\kp_0$ and $\Del$, the coupling constant $\kp_4$ controls the number of four-simplices, $N_4$, of the lattice configuration. In a given Monte Carlo  (MC) simulation it is convenient to keep $N_4$ (approximately) fixed. Thus $\kp_4$ will not play a direct role as a coupling constant. However, we will perform the MC simulations for various choices of $N_4$, and are interested in the limit $N_4 \to \infty$. For a given choice of $\kp_0$ and $\Del$ there is a critical value $\kp_4^c(\kp_0,\Del)$ of $\kp_4$, such that for $\kp_4 \to \kp_4^c(\kp_0,\Del)$ the expectation value of $N_4 \to \infty$ if we leave $N_4$ unconstrained in the MC simulations. In this way we are replacing the change in coupling constant $\kp_4$ with the change in $N_4$. In the $\kp_0,\Del$ coupling constant plane there are  phase transition lines\footnote{\label{foot-trans} Strictly speaking we only have genuine phase transitions in the limit $N_4 \to \infty$. However, for finite but large $N_4$, one observes a ``pseudo transition'' where order parameters change fast. By studying the location of these changes in the $\kp_0, \Del$ plane for a number of $N_4$'s, one obtains approximate locations of the phase transitions, which can be extrapolated to $N_4 \to \infty$. Such finite-size scaling studies will allow us to find the critical exponents related to the phase transitions, as we will discuss in detail below.}. These lines are boundaries for various regions in the bare coupling constant space, regions where typical geometries of the four-dimensional spacetime are very different. The phase diagram of the four-dimensional CDT model is shown in Fig. \ref{fig1} and we are here interested in the region named the ``de Sitter phase'' or the $C_{\rm dS}$ phase.

We will not discuss in any detail how to implement the numerical setup (again we refer to \cite{review1}), only mention that no background geometry is enforced in the computer simulations, but a time slicing, loosely corresponding to the proper time, and that the spatial {\it topology} of $S^3$ is also imposed on the lattice configurations for each leaf of the foliation. It should be emphasized that it does not mean that the computer $S^3$ configurations have the nice geometry of a round $S^3$ sphere embedded in $\mathbb{R}^4$: all possible (lattice) $S^3$ geometries accessible by the MC simulations. Thus the set up is as follows: each lattice geometry is defined by $N_4$ identical four-simplices where all links have length $a$ and at each time slice $t$ the three-spheres are constructed from $N_3(t)$ three-simplices (tetrahedra), also with link length $a$. Each such tetrahedron is shared by two four-simplices\footnote{\label{foot1}These four-simplices are denoted $(4,1)$ four-simplices because they have 4 vertices at the given time-slice and one vertex at the neighboring time-slice. To fill out the four-dimensional slab between two spatial time slices, one also needs the so-called $(3,2)$ four-simplices, where three and two vertices are placed on the neighboring time slices. We refer to \cite{review1} for a detailed discussion.}. The geometry is defined uniquely by the way the four-simplices are glued together. The Monte Carlo simulations change this gluing so that, in principle, one ``integrates'' over all geometries which can be constructed from these building blocks, preserving the topology and the foliation. For a fixed $N_4$ we can now measure $\la N_3(t)\ra$ and $\la N_3(t_1) N_3(t_2)\ra$ and thus observe some simple properties of our quantum universe, as well as construct an effective action that reproduces the ``observed'' results. In analogy with the continuum notation above, we introduce 
\beq\label{j14}
s_i = \frac{t_i}{N_4^{1/4}}, \quad n_3(s_i) = \frac{N_3(t_i)}{N_4^{3/4}}, \quad 
\Del s = \frac{1}{N_4^{1/4}},
\eeq 
and for given values of the bare coupling constants in the de Sitter phase we now have with high precision:
\beq\label{j12}
\la n_3(s) \ra = \frac{3}{4\om } \, \cos^3 \Big( \frac{ s}{\om}\Big), \quad 
\frac{\del}{\del_0} = \left(\frac{\om_0}{\om}\right)^{8/3},
\eeq
\beq\label{j13}
S_{\rm eff} = \frac{\sqrt{N_4}}{\G} \int_{-\pi \om/2}^{\pi \om/2} ds \; \left(
\frac{\dot{n}_3^2(s)}{n_3(s)} + \del \, n_3^{1/3}(s) \right), \quad
\int_{-\pi \om/2}^{\pi \om/2}  ds \, n_3(s) = 1,
\eeq   
where $S_{\rm eff} $ in \rf{j13} is the large $N_4$ limit of 
\beq\label{j15}
S_{\rm eff} = \frac{1}{\G} \sum_i  \left(
\frac{\big(N_3(t_{i+1}) -N_3(t_i)\big)^2}{ N_3(t_{i})\big)} + \del \, N_3^{1/3}(t_i) \right).
\eeq
We can clearly identify $\la n_3(s) \ra$ with the ``classical'' solution $n_3^{cl}(s)$ corresponding to the action \rf{j13}. $n_3^{cl}(s)$ will describe an ``elongated'' sphere if $\del > \del_0$, i.e. $\om < \om_0$ (the radius of the spatial three-spheres $n_3^{cl}(s)$ is multiplied by a factor $(\om_0/\om)^{4/3}$ compared to the three-spheres that enter the description of the ``round'' four-sphere with the same time extension). For the convenience of the reader we have in Appendix 1 listed the 
discrete variables and their continuum analogies.

\subsection{Renormalized versus bare coupling constants}

Before comparing CDT and FRG results let us recall how one moves between a lattice-regulated quantum field theory and the renormalized continuum quantum field theory defined from the lattice theory (assuming the continuum quantum field theory exists). This is essentially textbook stuff (see, e.g. \cite{munster}), but we feel it is useful for the following discussion to recall some details, which we have therefore collected in an Appendix.
Here we just emphasize the following: the lattice theory comes with a UV cut-off $a$, the length of the lattice links. We denote the coupling constants used to define the lattice theory by the bare coupling constants. The existence of a renormalized continuum theory usually requires that there exists a subset of bare lattice coupling constants, a so-called critical surface, where a correlation length $\xi$ is infinite, measured in lattice units. The renormalized coupling constants will depend on both the bare lattice coupling constants and the cut-off. By fine-tuning the bare coupling constants toward the critical surface in specific ways, one can achieve renormalized coupling constants that remain nontrivial in the limit as $a \to 0$. The correlation length $\xi$ will be defined in terms of the bare coupling constants, and let us for simplicity replace one of the bare coupling constants (typically the bare mass term) by $\xi$, and when we next talk about the bare coupling constants, we mean the remaining bare coupling constants. The critical surface will then be obtained by $\xi^{-1} \to 0$. Usually one can then relate the UV cut-off $a$ to $\xi$ by insisting that 
\beq\label{p3}
\xi \, a = \ell_{\rm phys}
\eeq
represents a physical length scale (the physical correlation length) that is kept fixed. If that is the case, taking $\xi^{-1} \to 0$ becomes equivalent to taking $a \to 0$.

\vspace{6pt}

\noindent
We can now take two different limits
\begin{enumerate} 
\item[1)] One can keep the bare lattice coupling constants fixed and take the correlation length $\xi$ to infinity. In that limit the renormalized coupling constants, which depend on both the bare coupling constants and the correlation length, will flow to an IR fixed point of the renormalized theory.
\item[2)] Alternatively, one can keep the renormalized
coupling constants fixed when taking the correlation length to infinity. This will in general only be possible if we adjust the bare coupling constants, and in this way they become functions of the correlation length (and the fixed renormalized coupling constants) and when $\xi \to \infty$ the bare lattice coupling constants will flow to a lattice UV fixed point.
\end{enumerate}

The scenario above assumes that the lattice considered is infinite since it is $\xi = \infty$ on a critical surface. In lattice field theory MC simulations will be instead performed on a finite lattice with $N = L^d$ number of lattice sites where $L$ denotes the linear extension of the lattice and $d$ the dimension of spacetime. $\xi = \infty$ is then replaced by $\xi \propto N^{1/d}$, since the correlation length cannot exceed the linear dimension of the lattice. In lattice field theory this leads to the concept of a pseudo-critical surface in the coupling constant space, a surface that will approach the critical surface for $N \to \infty$, and to the theory of finite-size scaling, where scaling behavior of observables when one approaches the critical surface can be extracted by studying the behavior of the observables at the pseudo-critical points, as a function of $N$ (as discussed in the Appendix 2). In the above discussion, the correlation length $\xi$ is related to some matter fields living on the lattice sites, and its only relation to the lattice volume $N$ is by various finite-size effects, for example, $\xi_{\rm max} \propto N^{1/d}$. In quantum gravity the situation is different since the volume of spacetime $V_d$ (or the lattice spacetime volume $N$) is a dynamical variable, and the way we have encountered finite-size scaling in \rf{j14}-\rf{j15} for variables $\la N_3(t) \ra_{N_4}$ and $\la N_3(t) N_3(t') \ra_{N_4}$ is slightly different: $N_4^{1/4}$ {\it is} (essentially) the correlation length.  An appropriate question is then: the correlation length of what? The answer is: the correlation between spacetime points separated by a geodesic distance $n$ in the triangulation. This can be seen explicitly in two-dimensional quantum gravity where the various 2d lattice models of quantum gravity can be solved analytically \cite{aw,book,aletal}. Let $\mu$ denote the lattice cosmological constant, $\mu_c$ its critical value, and $N_2$ the number of triangles in a 2d triangulation. One has 
\beq\label{jd2a}
\la N_2 \ra \propto \frac{1}{\mu_c \mi \mu}, 
\qquad P_\mu(n) \propto \e^{-n/\la N_2 \ra^{1/d_h} } \quad \text{for} \ n > \la N_2\ra^{1/d_h}. 
\eeq
Here $d_h$ denotes the Hausdorff dimension of the 2d lattice gravity model and $P_\mu(n)$ the probability for two marked points to be separated by a geodesic distance $n$ in the statistical ensemble of triangulations with cosmological constant $\mu$. It can be seen that $ \la N_2\ra^{1/d_h}$ is the correlation length and goes to infinity as $\mu \to \mu_c$, the critical value of the cosmological coupling constant. The relation between the two-point function where the cosmological constant $\mu$ is fixed and the one where its conjugate variable, the two-volume $N_2$, is fixed, is given by a Laplace transformation
\beq\label{jd2b}
P_\mu(n) = \sum_{N_2} \e^{-(\mu-\mu_c) N_2} P_{N_2} (n), \quad \textrm{i.e.} \quad
P_{N_2}(n)  \propto \frac{1}{N_2^{1/d_h}}  F\Big( \frac{n}{N_2^{1/d_h}} \Big)
\eeq
for $N_2$ sufficiently large, so that the discretization effects are small. The point we want to emphasize is the appearance of the scaling function $F(x)$, $x =n/N_2^{1/d_h}$ for a fixed, large $N_2$. It is a consequence of the divergence of the correlation length $ \la N_2\ra^{1/d_h}$ for $\mu \to \mu_c$, and for sufficiently large $N_2$ we can identify $N_2 \approx  \la N_2\ra$ and $N_2^{1/d_h}$ with the correlation length. In 4d CDT we have a finite volume, since $N_4$ is always finite in the computer simulations. However, we observe that finite-size scaling works very well \cite{hausdorff4,agjl1,cdt-finitesize}, and the measurements of $\la N_3(t) \ra_{N_4}$ and $\la N_3(t) N_3(t')\ra_{N_4}$ lead to finite-size scaling functions $F(x)$ as shown, e.g., in \rf{j12}-\rf{j13}. In fact, as discussed in detail in \cite{reviewfrontier}, the correlation function $\la N_3(t) N_3(t')\ra_{N_4}$ is precisely the two-point correlator integrated over two spatial hyperplanes separated by a geodesic distance proportional to $|t-t'|$, a setup closely analogous to the setup used to measure the correlation length of a scalar field in a $\phi^4$ theory on a lattice
\cite{munster}.

Since the Hausdorff dimension of four-dimensional CDT is four \cite{hausdorff4}, it is then natural to assume that we have a correlation length $\xi \propto N_4^{1/4}$ and in this case the relation between the continuum four-volume $V_4$ and the lattice volume $N_4$ becomes similar to \rf{p3}
\beq\label{p4}
N_4 a^4 \propto V_4, \quad {\rm i.e.} \quad 
\xi a \propto \ell_{\rm phys} = V_4^{1/4},
\eeq
We want to apply 1) and 2) to CDT, with $\xi$ being $N_4^{1/4}$ and the bare couplings being $\kp_0$ and $\Del$, and the critical surface being $N_4 \to \infty$. For fixed $\kp_0$ and $\Del$ we then expect the renormalized coupling constants (which we have not yet defined) to flow to an IR fixed point. Similarly, keeping the renormalized coupling constants fixed, the bare coupling constants $\kp_0,\Del$ should flow to a lattice UV fixed point (provided it exists). We will discuss this in detail below.

\subsection{The infrared limit}

We want to compare the lattice results \rf{j12} and \rf{j13} with the FRG results \rf{j8} and \rf{j8a}, where in formulas \rf{j12} and \rf{j13}  the lattice coupling constants $\G$ and $\del$ are functions of the bare lattice coupling constants $\kp_0$ and $\Del$.\footnote{A first such comparison was done in \cite{cdt-rgf}. However, at that time the so-called bifurcation phase $C_b$ had not yet been discovered. It was viewed as part of phase $C_{\rm dS}$.}
Let us for the moment ignore that $\del \neq \del_0$.
Then it is natural to identify
\beq\label{j16}
\frac{\sqrt{N_4}}{\G(\kp_0,\Del,N_4)} = \frac{\sqrt{V_4(k)}}{24 \pi G_k} \simeq \frac{1}{1.63\, \lam_k g_k}.
\eeq
The equation provides us with a mapping from the bare lattice couplings $\kp_0,\Del,N_4$ to the renormalized running coupling constant $\lam_kg_k$ that we then view as a renormalized lattice constant combination.
As long as we keep the bare lattice coupling constants $\kp_0$ and $\Del$ fixed, $\G$ will essentially stay fixed for sufficiently large $N_4$.
As discussed above we view $N_4^{1/4}$ as proportional to a correlation length $\xi$.
Approaching the critical surface $N_4^{-1} = 0$ should thus lead to a flow of the renormalized coupling combination to an IR fixed point,
and we see that this IR fixed point should be one where $\lam_kg_k \to 0$.
Let us now check if this agrees with the IR limit obtained from FRG for $k \to 0$.
  
If we assume that both $\lam_k$ and $g_k$ are small, then we are close the Gaussian fixed point of the renormalization group flow and lowest order perturbation theory yields (e.g. see the linear approximation to Eq.\ (74) in \cite{review3}):
 \beq\label{ju}
 g_k = g_{k_0} \frac{k^2}{k_0^2}, \quad
 \lam_k = \Big(\lam_{k_0} \mi \frac{g_{k_0}}{8\pi} \Big)\, \frac{k_0^2}{k^2} + 
  \frac{g_{k_0}}{8\pi} \,\frac{k^2}{k_0^2}, \quad k \approx k_0, ~~ g_{k_0},\lam_{k_0} \ll 1.
  \eeq
When $k \to 0$ one has $\lam_k \to \infty$ unless $g_{k_0} = 8\pi \, \lam_{k_0}$, in which case we start out precisely at the unique renormalization trajectory that leads to the Gaussian fixed point. Unless that is the case, lowest order perturbation theory will become invalid for $k \to 0$, since $\lam_k \to \infty$. In most FRG approaches, the $\beta$-functions become singular when $\lam_k =1/2$, and this happens for finite $k$. An improved FRG treatment (that goes far beyond the simple Ansatz \rf{j1}) has led to the following limit for $k \to 0$ 
 \cite{pawlowski}:
 \beq\label{ju1}
 g_k \propto \frac{k^2}{k_0^2}, \quad
 \lam_k  \propto \frac{k_0^2}{k^2} , \quad {\rm for}\quad k \to 0,
 \eeq
despite the fact that $\lam_k \to \infty$ for $k \to 0$. This limit then serves as an IR fixed point in the sense that it corresponds to finite values of the dimensionful gravitational and the cosmological constants: $\Lam_{k=0} \approx \Lam_{k_0}$ and $G_{k=0} \approx G_{k_0}$ in the limit $k \to 0$.
Eqs.\ \rf{ju1} also imply that $g_k \lam_k = G_k \Lam_k > 0$ for $k \to 0$ and is therefore not compatible with the large $N_4$ limit of \rf{j16}. If one wants an infrared limit like \rf{ju1} one has to change the bare coupling constants of the lattice theory such that $\G(\kp_0,\Del) \propto \sqrt{N_4}$ in the relation \rf{j16}, a change that should not be needed in order to approach the IR fixed point, as discussed in  Appendix 2.
However, we will discuss the $\G(\kp_0,\Del) \propto \sqrt{N_4}$ behavior later in detail,
since it is the same behavior that is needed to reach a UV fixed point.
 
However, in a recent paper \cite{frank-w} the $k \to 0$ limit has been studied using FRG for time-foliated spacetimes, a setup that is closer to our CDT approach and a new IR fixed point was found:
\beq\label{ju3}
 (g_k,\lam_k) \to \left(0, \oh\right) \quad {\rm for} \quad k \to 0.
\eeq
More precisely it was found that
\beq\label{ju4}
 g_k \propto \frac{k^4}{k_0^4}, \quad \lam_k \mi \oh \propto \frac{k}{k_0}, \quad {\rm for} \quad 
 k \to 0.
\eeq
Except for the unique path from the UV fixed point to the Gaussian fixed point, the paths from the UV fixed point (corresponding to $k =\infty$) will all converge to this IR fixed point for $k \to 0$. When approaching the fixed point, we have $g_k \lam_k \to 0$.
This is then compatible with the CDT limit for $N_4 \to \infty$ and $\kp_0,\Del$ fixed.
This IR fixed point is different from the Gaussian fixed point since the approach to the Gaussian fixed point can be parametrized by a classical gravitational coupling constant $g_k/k^2 = G_k \to G_0$, while $\Lam_k \to 0$ as $ k^2$. For the IR fixed point we have $G_k \propto k^2 \to 0$, and also $\Lam_k \propto k^2 \to 0$. As we will now argue, this different scaling has implications for the scaling of the lattice cut-off $a$.

First we remark that \rf{p4} and \rf{j16} lead to further identification
\beq\label{js1}
\G a^2 \propto G_k .
\eeq
In the case of the Gaussian fixed point, we can write $G_k = G_0$ for small $k$
and define the Planck length as $l_p = \sqrt{G_0}$. Thus \rf{js1} becomes
\beq\label{js2}
a \propto \frac{l_p}{\sqrt{\G} }  \quad  {\rm for} \quad N_4 \to \infty.
\eeq

Since for fixed bare lattice coupling constants $\kp_0,\Del$, the measured $\G(\kp_0,\Del,N_4)$ goes to a constant for $N_4 \to \infty$, we see that the UV cut-off $a$ is of the order of the Planck length $l_p$, and it does not go to zero at the Gaussian fixed point.
In the interior of the $C_{\rm dS}$ phase where $\G$ is bounded we reach the interesting conclusion
that we cannot take the cut-off (much) below the Planck scale when approaching the Gaussian fixed point.
We thus have a situation where \rf{p3} is not really valid in the sense that when approaching an IR fixed point in the renormalized coupling constants, we cannot maintain $\ell_{\rm phys}$ as a constant.
Eq. \rf{p4} is of course still valid, but $\ell_{\rm phys} = V_4^{1/4}(k) \to \infty$ for $k \to 0$ since $V_4(k) \propto 1/\Lam_k^2$ and $\Lam_k \to 0$ at the Gaussian fixed point, so in this case \rf{p4} can be satisfied for $a = \mathrm{const.}$ and $\xi \to \infty$.

The situation is different when approaching the IR fixed point \rf{ju4}, since we in that case have $G_k \propto k^2$, and we obtain from \rf{js1}
\beq\label{js3}
 a \propto \sqrt{\frac{g_{k_0}}{\G}} \, \frac{k}{k_0^2}, \qquad \frac{g_{k_0}}{k_0^4} \approx \mathrm{const.} \quad
 {\rm for} \quad k_0 \to 0.
\eeq

Thus, $a \to 0$ for $k \to 0$. So in this case, we obtain a simple continuum limit at the IR fixed point.
However, the continuum theory determined by the FRG is in some sense a trivial theory since both $G_0$ and $\Lam_0$ are zero.

For future reference let us note that close to the Gaussian fixed point \rf{ju} and the IR fixed point \rf{ju4} we have what can be considered as the leading contribution to the renormaliztion group equation for 
 $\lam_k g_k$ for $k \to 0$:
 \beq\label{js4}
 k \frac{ d }{dk}(\lam_k g_k) = 4 (\lam_k g_k) + O( k^5)\quad
 {\rm for} \quad k \to 0.
 \eeq

\subsection{The ultraviolet limit}
 
We are of course more interested in the UV fixed point.
Following 2) (and the discussion in  Appendix 2) we can locate a lattice UV fixed point by keeping
the renormalized couplings fixed while approaching the critical surface, i.e. while following a path in the
$\kp_0,\Del$ coupling constant space where the correlation length $N_4^{1/4} \to \infty$.
The relation \rf{j16} allows us to view $\lam_k g_k$ as a combination of renormalized coupling constants also on the lattice.
In the FRG approach this renormalized coupling is a running coupling,
taking values between its IR and UV values.\footnote{In most 
FRG approaches one has a coupled set of RG equations for 
$\lam_k$ and $g_k$. However, in \cite{kawai} it was possible to derive a closed
FRG equation for $\lam_kg_k$ and the $\beta$-function for $\lam_kg_k$ has both an IR fixed point, $\lam_0g_0 =0$,
for $k \to 0$, and an UV fixed point $\lam^*g^*$ for $k \to \infty$.}
In relation \rf{j16}, picking any value of $\kp_0,\Del,N_4$
will then (in principle) correspond to a value of $\lam_kg_k$
between its IR value and its UV value.
We are then instructed to take $N_4 \to \infty$ while at the same time changing $\kp_0$ and $\Del$ so that $\lam_k g_k$ stays fixed.
The path followed in the $\kp_0,\Del$ plane will then lead us to the lattice UV point, provided that it exists.
This flow in the lattice coupling constants is not directly related to $k$-flow of the renormalized coupling.
As explained in  Appendix 2 one should be able to reconstruct the flow of the renormalized $\lam_k g_k$
(as a function of the correlation length $\xi = N_4^{1/4}$) by considering the whole family of flows to the UV fixed point,
starting from different $\kp_0,\Del,N_4$.
However, here we will only be interested in locating the UV fixed point and finding the critical exponents, as described in Appendix 2.

According to the discussion above, we have to require
\beq\label{j17}
\frac{\sqrt{N_4}}{\G(\kp_0,\Del,N_4)} = {\rm constant}
\quad {\rm for} \quad N_4 \to \infty,
\eeq
in order to approach a lattice UV fixed point.
For fixed $\kp_0,\Del$ in the interior of the $C_{\rm dS}$ phase $\G(\kp_0,\Del,N_4)$ stays finite for $N_4 \to \infty$.
Thus, a path $N_4 \to \big(\kp_0(N_4),\Del(N_4)\big)$ in the $\kp_0,\Del$ coupling constant plane
where $\G\big(\kp_0(N_4),\Del(N_4),N_4\big) \to \infty$ for $N_4 \to \infty$
will take us to the phase boundaries of the $C_{\rm dS}$ region.
There are three phase transition lines,
the $C_b\mi C_{\rm dS}$ transition line, the $B \mi C_{\rm dS}$ transition line, and the $A \mi C_{\rm dS}$ transition line.
As we will discuss in the numerical Section below, only at the $A \mi C_{\rm dS}$ transition line $\G$ diverges.
We conclude that a putative UV fixed point has to be located along this line and could even be a UV transition line.

In earlier work we have classified the  $A \mi C_{\rm dS}$ 
transition line as a first order transition line  \cite{cdt-finitesize,Ambjorn:2019pkp}. However, this has been based on the study of certain local order parameters. The usual paradigm for a first order transition associated with a matter field is that when the lattice size of system is infinite ($N_4$ infinite in our case) the correlation length $\xi$ associated with the matter field will stay finite even when we approach the phase transition line. This situation cannot be realized when we associate the correlation length with $N_4^{1/4}$ and take $N_4 \to \infty$. The situation is thus more complicated for phase transition lines bordering the $C_{\rm dS}$ phase, and we will address this further below.

Before turning to the numerical results, we have to deal with the fact
that in general $\del$ in \rf{j13} will be different from $\del_0$ in \rf{j6}.

\subsection{Dealing with $\pmb{\del \neq \del_0}$ }

We now need to be more precise when comparing CDT lattice results with the FRG results. Let us first note that the signs of the actions in Eqs.\ \rf{j8} and \rf{j13} are opposite. Although CDT rotates to an Euclidean signature , the requirement of convergence of the path integral (i.e. convergence of the sum of triangulations with the appropriate Boltzmann weight) also modifies the way the conformal factor will appear
(see \cite{review1} for a discussion), resulting in the sign change. In fact the minisuperspace action \rf{j13} is precisely the Hartle-Hawking minisuperspace action after the rotation of the conformal factor. In this sense the sign difference is natural. Secondly, the measured values of $\om$ in the lattice simulations are in general different from the value $\om_0$ dictated by GR. In \cite{cdt-rgf} we tried to fit it into the framework of Horava-Lifshitz gravity (HLG) \cite{horava}\footnote{It is natural to try to fit CDT into the framework of HLG since in both theories time has a special role compared to space. This has indeed been attempted, but not very successfully. The asymmetry between space and time in HLG is caused by adding higher derivative spatial terms to the action. Such terms are not added to the CDT action. In fact the asymmetry observed in CDT  does not scale in the way predicted by HLG.}. Here we will be more conservative and try to fit our data to GR represented by \rf{j6}. Since we explicitly break the symmetry between space and time in our lattice regularization, we also have the freedom to scale space-like links and time-like links differently in order to obtain continuum results compatible with the spacetime symmetry present in GR. Denote the length of the time-like links by $a_t$ and the length of the space-like links by $a_s \equiv a$. The continuum three-volume of  a spatial slice at time $t_i$, consisting of $N_3(t_i)$ tetrahedra will then be $V_3(t_i) \propto N_3(t_i) a^3$. Similarly, the continuum four-volumes of $N_4$ four-simplices will be $V_4 \propto N_4 a_t a^3$. Strictly speaking the situation is somewhat more complicated for the four-simplices\footnote{To be more precise, the change of $V_4$ is as follows (see \cite{review1} for a detailed discussion of the change of the volumes of the two kind of  four-simplices mentioned in footnote \ref{foot1}, when the relative length of $a_s$ and $a_t$ are  $a_t^2 = \alpha a_s^2$): the four-volume of a simplex with side length $a$ changes from $\sqrt{5}/96 \cdot a^4$ to 
\beq\label{jf1}
V_4^{(4,1)}= \frac{a_t a^3 _s}{96}
\sqrt{8 \mi \frac{3}{\alpha}},
\qquad  V_4^{(3,2)} = \frac{\a_t a_s^3}{96} 
\sqrt{12 \mi \frac{7}{\alpha}}
\eeq 
However, we will mainly be interested in the  behavior in regions of coupling constant space where $\om/\om_0$ is significantly smaller than 1, i.e.\ according to \rf{jk10} $\alpha$ much larger than $1$,
and then \rf{jw2} will be good enough for our discussion.}.
However, for notational simplicity we will simply write
\beq\label{jw2}
V_4 = N_4 a_t a^3, \quad V_3 =  N_3 a^3.
\eeq
Next, introducing $V_3$ and $V_4$ we can now write \rf{j15} as 
\bea\label{jk6}
S &=& \frac{1}{\G} \sum_i \left(\frac{( N_3(t_i\plu a_t) \mi N_3(t_i))^2}{N_3(t_i)} + 
\del \;N_3^{1/3}(t_i)\right)\quad ~~( t_i \equiv a_t  i) \\
&=& \frac{a_t}{a^3\G} \sum_i  a_t \left( 
\frac{( V_3(t_i+a_t) \mi V_3(t_i))^2/a_t^2}{V_3(t_i)} + 
\frac{a^2}{a_t^2} \,{\del}\; V_3^{1/3}(t_i)\right),
\label{jk7}\\
&\to& \frac{1}{ 24 \pi G} \int dt \left( \frac{\dot{V}_3^2}{V_3} + \tilde{\del }\; V_3^{1/3} \right),
\quad \tilde{\del} = \frac{a^2}{a_t^2} \,\del, ~~ 24\pi G = \frac{a^3}{a_t}\,\G, ~~
\hspace{0.8cm}
\label{jk8}
\eea 
and 
\beq\label{jk9}
\sum_i N_3(i) = N_4 \to \int dt \,V_3(t) = V_4, \qquad 
V_4 = a_t a^3 N_4.
\eeq
where $\tilde{\del}$ and  $\om$ are related as in \rf{j12}: $\tilde{\del}\, \om^{8/3} =  \del_0 \om_0^{8/3}$. If $\om \neq \om_0$ then  $\tilde{\del} \neq \del_0$. On the lattice  the ``deformed'' spheres arise because there are ``too many'' $N_3$'s compared to the time extension $N_t = \om N_4^{1/4}$ measured. We can fix that by adjusting $a_t$ such that the continuum, physical time length $N_t a_t$ agrees with the spatial extention $N_3^{1/3} a$,
where we write $N_4 = N_t N_3$.
From these equations we find 
\beq\label{jk10}
a_t = \left(\frac{\om_0}{\om}\right)^{4/3} a,
\eeq
where we have adjusted the various constants entering in the estimate such that $a_t = a$ if $\om = \om_0$, the value for the geometry of a round $S^4$. From Eq.\ \rf{jk8} it then also follows that $\tilde{\del} = \del_0$ for the round $S^4$ and thus this choice of $a_t$ leads to an action $S$ given in \rf{j8} that we can identify with the FRG effective action for some value of the scale parameter
$k$. 

So given computer data $N_4,\om,\G$ we can associate a corresponding continuum, round  $S^4$ with four-volume $V_4$ and gravitation constant $G$ via:
\beq\label{jh4}
(N_4,\om,\G)  \to (V_4(k),\om_0,G_k), 
\eeq
 where
 \beq\label{jk11}
 V_4(k) = \left( \frac{\om_0}{\om} \right)^{4/3} N_4 a^4, \quad 
 24 \pi G_k =  \left( \frac{\om}{\om_0} \right)^{4/3} \G \, a^2
 \eeq
 and in particular
 \beq\label{jh2a}
\frac{\sqrt{N_4}}{\G} = 
\frac{\om^2}{\om_0^2}\; \frac{\sqrt{V_4(k)}}{24\pi G_k} \qquad 
{\rm or} \qquad 
\frac{\om^2 \G(\kp_0,\Del,N_4)}{\om^2_0 \sqrt{N_4}} \simeq 1.63\, \lam_k g_k.
\eeq

As mentioned, the only way we can influence the values of $\om$ and $\G$ is by choosing the bare lattice coupling constants $\kp_0$ and $\Del$. Both $\om$ and $\G$ depend only weakly on $N_4$ for sufficiently large $N_4$. We find numerically for all the bare lattice coupling constants in phase $C_{\rm dS}$ that $\om^2 \G$ is bounded except close to the $A \mi C_{\rm dS}$ transition. In particular, at the $C_b \mi C_{\rm dS}$ transition line, which is a second order transition line \cite{Coumbe2016,Coumbe2017}, it seems that $\om^2 \G$ stays bounded and different from zero. There might be huge finite-size effects at this second order phase transition line, since we have good reasons to believe that $\om = 0$ deep into the $C_b$ phase, but we do not observe any dip of $\om$ close to the transition line. Similarly, there are large finite-size effects at the little stretch of the $B \mi C_{\rm dS}$ transition, which is probably also a second order phase transition line.
On the other hand, the situation at the $A \mi C_{\rm dS}$ transition is relatively clear: $\G \to \infty$ and $\om \to 0$ and, most importantly,
$\om^2 \G \to \infty$  (in the limit $N_4 \to \infty$) when approaching the transition from the $C_{\rm dS}$ side.

\section{The putative UV limit}

A necessary condition for obtaining a UV limit from the lattice theory is that we, starting from a bare coupling constant point $(\kp_0,\Del,N_4(0))$ associated to phase $C_{\rm dS}$ \footnote{By that we mean that keeping $\kp_0$ and $\Del$ fixed and taking $N_4 \to \infty$ we end up in phase $C_{\rm dS}$, what is strictly speaking only defined for $N_4 = \infty$.} can find a path $(\kp_0(N_4),\Del(N_4),N_4)$ such that 
\beq\label{jo1}
\frac{\om^2 (\kp_0(N_4),\Del(N_4),N_4)
\G(\kp_0(N_4),\Del(N_4),N_4)}{\om^2_0 \sqrt{N_4}} \simeq 1.63\, \lam_k g_k,
\eeq
for some value of $k$ and all the way to $N_4 = \infty$, as already mentioned above (Eq.\ \rf{jh2a}). As discussed in the numerical results section below, the observed dependence on $\Del$ is weak in the region of interest, and for notational simplicity, we will omit references to $\Del$ in the following.

As discussed in  Appendix 2 we assume that at the critical surface, i.e.\ the $N_4 = \infty$ surface, we have the following behavior of $\G(\kp_0)$ and $\om(\kp_0)$ close to the critical point $\kp_0^\UV$ of the $A\mi C_{\rm dS}$ 
transition:
\beq\label{jo2}
\G(\kp_0) \propto \frac{1}{| \kp_0^{\UV} \mi \kp_0|^\alpha}, 
\quad \om(\kp_0) \propto | \kp_0^{\UV} \mi \kp_0|^\beta, \quad 
 \om^2(\kp_0)\G(\kp_0) \propto \frac{1}{| \kp_0^{\UV} \mi \kp_0|^{\alpha-2\beta}},
 \eeq
We further assume that for a finite $N_4$ there is a pseudo-critical point $\kp_0^{\UV} (N_4) < \kp_0^{\UV}$ where $\om^2(\kp_0,N_4)\G(\kp_0,N_4)$ has a maximum for fixed $N_4$, and that this pseudo-critical point approaches $\kp_0^{\UV}$ for $N_4 \to \infty$ as 
\beq\label{jo3}
\kp^{\UV}_0(N_4) = \kp_0^{\UV} -\frac{c}{N_4^{1/4\nu_{\UV}}} 
\qquad \Big( {\rm i.e.} 
\quad \xi \propto \frac{1}{|\kp_0^{\UV} \mi \kp_0^{\UV}(\xi)|^{\nu_{\UV}}} 
\Big).
\eeq
This implies that 
\beq\label{jo4}
\G(\kp_0^{\UV}(N_4) ) \propto N_4^{\alpha/4\nu_{\UV}}, 
\quad \om(\kp_0^{\UV}(N_4)) \propto N_4^{-\beta/4\nu_{\UV}}, 
 \eeq
 as well as 
 \beq\label{jo5}
 \om^2(\kp_0^{\UV}(N_4))\G(\kp_0^{\UV}(N_4)) \propto 
 N_4^{(\alpha-2\beta)/4\nu_{\UV}}.
\eeq
From Eq.\ \rf{jo1} it follows that we have to have 
\beq\label{jo6}
\alpha - 2 \beta \geq 2 \nu_{\UV}
\eeq
and if that is the case the following path in the bare lattice coupling constant 
space will lead us to the putative UV fixed point while keeping $\lam_k g_k$ fixed:
\beq\label{jo7}
\kp_0(N_4) = \kp_0^{\UV} - \frac{c}{N_4^{1/2(\alpha - 2 \beta)}}
\eeq 
Following the discussion in  Appendix 2, only with 
$\alpha$ there replaced by $\alpha -2\beta$, Eq.\ \rf{add3} in  Appendix 2 suggests that we should associate a critical 
exponent
\beq\label{add4}
\theta_{\kp_0^{\UV}} = \frac{2}{\alpha - 2 \beta}
\eeq
to the putative UV fixed point. Presently, it is unclear to us if this can be directly compared to the corresponding critical exponent obtained using the FRG approach. The main problem is that we do not know the precise connection between our correlation length $\xi \propto N_4^{1/4}$ used to derive $\theta_{\kp_0^{\UV}}$ and the scale factor $k$ used in the FRG.

The MC simulations discussed in next Section indicate that we have 
\beq\label{jo8}
\frac{\beta}{4\nu_{\UV}} \approx 0.24\pm 0.02, \quad \frac{\alpha}{4 \nu_{\UV}} 
\approx 1.04\pm 0.02, \quad \nu_{\UV} \approx 0.21\pm 0.03 , 
\eeq
where we have taken the average values measured using two different methods.
Clearly it would be very desirable if we could measure the exponents with larger precision, but it will require considerably larger computer simulations.
In the following discussion we will assume that the actual exponents satisfy
\beq\label{jo9}
\frac{\alpha - 2 \beta}{4\nu_{\UV}} > \oh , \qquad \frac{\beta}{4\nu_{\UV}} < \frac{1}{4},
\eeq
and let us understand better the relation between 
the FRG flow and what we observe on the lattice. From \rf{jk11} we find, using that
$\om \propto N_4^{-\beta/4\nu_{\UV}}$ and $V_4(k) \propto \Lam_k^{-2}$, that 
\beq\label{jo9a}
a \propto \frac{1}{ \sqrt{\Lam_k}}\; N_4^{-\frac{1}{4}(1 + \frac{\beta}{3\nu_{\UV}})},
\quad {\rm i.e.} \quad a  \propto \frac{1}{ k}\, N_4^{-\frac{1}{4}(1 + \frac{\beta}{3\nu_{\UV}})} 
\quad {\rm for} \quad k \to \infty.
\eeq
Thus the lattice spacing $a$ scales to zero for a fixed value of $k$ when we approach the critical surface $N_4 = \infty$. From Eqs. \rf{jk11} and \rf{jo9a} it follows that also the $a_t$ scales to zero when approaching the  critical surface, but slower:
\beq\label{jo10} 
a_t \propto \frac{1}{\sqrt{\Lam_k}} \; N_4^{-\frac{1}{4}( 1- \frac{\beta}{\nu_{\UV}})}
\quad {\rm i.e.} \quad a_t  \propto \frac{1}{ k}\, N_4^{-\frac{1}{4}(1 - \frac{\beta}{\nu_{\UV}})} 
\quad {\rm for} \quad k \to \infty.
\eeq
This slower decrease of $a_t$ is a reflection of the fact that we, when approaching the $A\mi C_{\rm dS}$ transition line, have to rescale our increasingly ``elongated'' lattice four-spheres in order to match the round four-spheres of the FRG. Thus, under the assumptions stated in \rf{jo9} the $A \mi C_{\rm dS}$ phase transition point $\kp^{\UV}_0$ can serve as a UV fixed point and approaching the fixed point we have a well-defined mapping of our deformed four-spheres to the four-spheres predicted by the simplest FRG equations all the way to the lattice UV fixed point.
Recall again to the reader that this flow occurs for a fixed value of $\lam_k g_k$.
Thus, it is not directly related to the FRG flow parameter $k$ and the flow of $\lam_k g_k$ to its UV fixed point.
However, the larger $k$ the smaller the lattice spacing $a$ and we see that when $k \to \infty$ then $a \to 0$ as $1/k$, simply reflecting that $V_4(k) \to 0 $ as $1/k^4$. Let us emphasize that the real content 
of Eqs.\ \rf{jo9a} and \rf{jo10} is not this relation between $a$ and $k$,
since $k$ is kept fixed when we approach the UV fixed point, but that 
$a \to 0$ when the correlation length $N_4^{1/4} \to \infty$. This is 
why we might be able to take the ``continuum'' limit at the UV fixed point.

Let us finally note that the UV critical exponent $\theta_{\kp_0^{\UV}}$, defined by Eq. \rf{add4}, given the numerical values \rf{jo8}, will be
\beq\label{add5}
\theta_{\kp_0^{\UV}} = 4 \pm 1.
\eeq
This value is larger than the one reported in \cite{kevin} and much larger than the one reported in \cite{kawai}, but as already emphasized, we do not know the precise relation between the FRG $k$ and our $\xi$. Unfortunately, Eqs. \rf{jo9a} and \rf{jo10} do not help us, since the lattice spacing $a$ also appears.\footnote{A further discussion of the relation between the critical exponents will appear in \cite{kevin-renata}.}

If Eqs. \rf{jo9} are \textit{not} satisfied it is difficult to associate a UV fixed point to the $A \mi C_{\rm dS}$ phase transition,
at least following the philosophy used in this article.
It is of course unfortunate that the numerical results do not allow us to decide whether or not a UV fixed point is favored.

\section{Numerical results}

\begin{figure}
    \centering
    \includegraphics[width=0.45\linewidth]{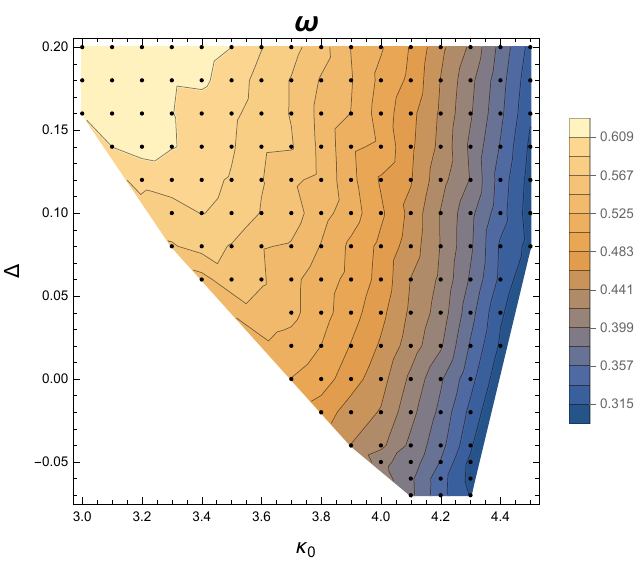}
    \includegraphics[width=0.45\linewidth]{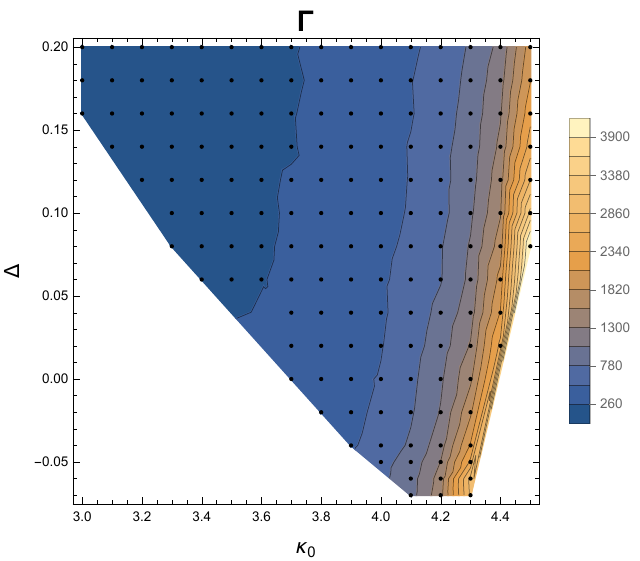}
     \includegraphics[width=0.45\linewidth]{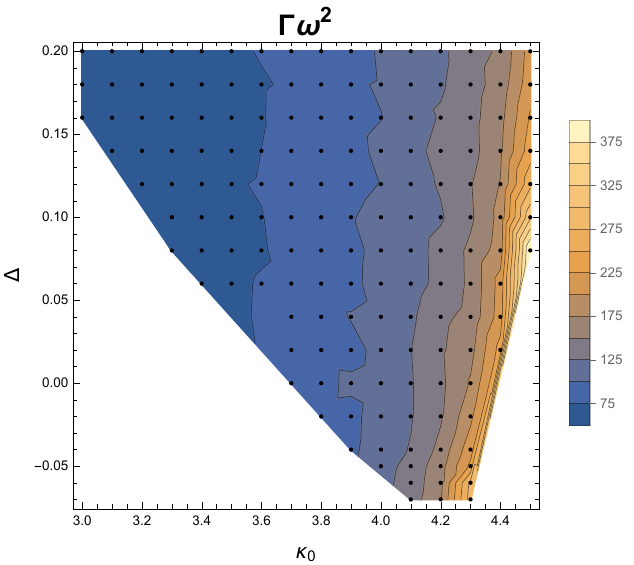}
    \caption{Contour plots of $\omega$ (left), $\Gamma$ (right) and $\Gamma\omega^2$ (bottom) in the function of the CDT coupling constants $\kappa_0, \Delta$. Points where actual measurements were done are denoted as black dots in the plots.}
    \label{fig3}
\end{figure}

\begin{figure}
    \centering
    \includegraphics[width=0.45\linewidth]{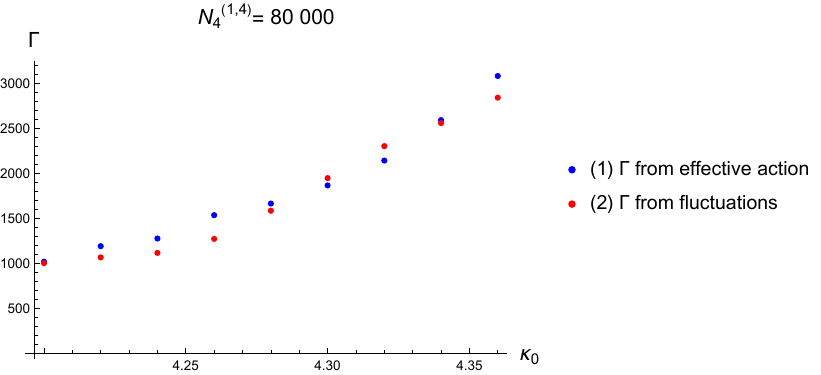}
    \includegraphics[width=0.45\linewidth]{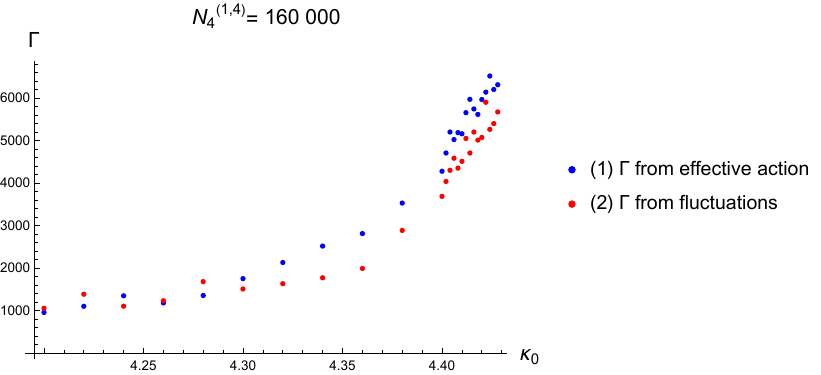}
    \includegraphics[width=0.45\linewidth]{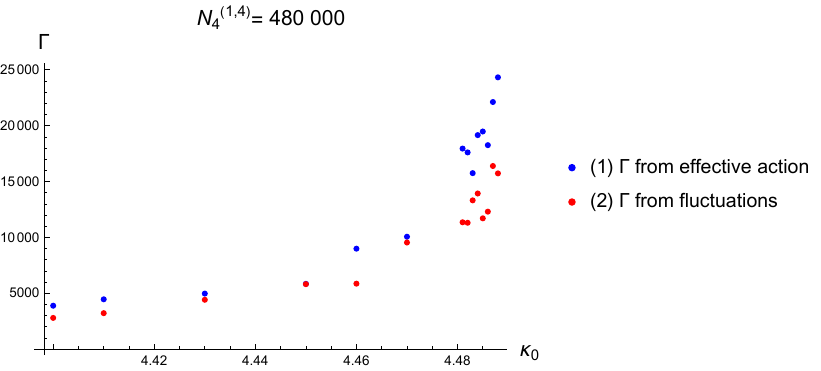}
    \includegraphics[width=0.45\linewidth]{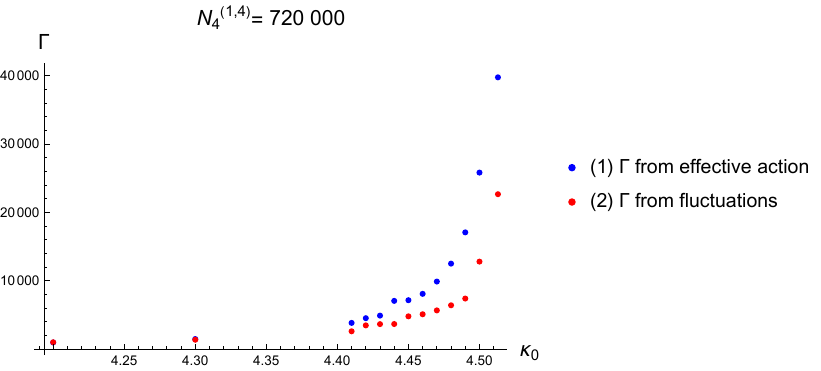}
    \caption{Values of $\Gamma$ computed using two different methods discussed in the text, measured for $\Delta=0$ and a choice of $\kappa_0$ near the $A-C_{\rm dS}$ transition for $N_4^{(4,1)} = 80\, 000, 160\, 000, 480\, 000, 720\, 000$. }
\label{fig2}
\end{figure}

In this Section, we present results of the numerical Monte Carlo (MC) simulations performed inside the de Sitter phase $C_{\rm dS}$ in a region of $\kappa_0, \Delta$ coupling constant space close to the three phase boundaries discussed above; see also Figures~\ref{fig1} and~\ref{fig3}.\footnote{One should note that in the previous study of this region of the parameter space \cite{cdt-rgf} the existence of the bifurcation phase $C_b$ was not known,
and therefore one could not analyze all phase transition lines in a proper way.
Also, because of much lower computing power, one was only able to simulate quite small triangulations (up to $N_4^{(4,1)} =  40\,000$).
Now, thanks to better computer resources and improved MC algorithms, we were able to simulate much larger systems (up to $N_4^{(4,1)} = 720\,000$)
and thus considerably reduce finite-size effects.}
As already explained, the measured values of $\omega$ and $\Gamma$ are both functions of $\kappa_0$ and $\Delta$,
but for sufficiently large systems, do not depend on $N_4$ (except very close to the phase boundaries, which also shift with $N_4$).
Therefore, as a starting point, to check the dependence one can perform MC simulations for fixed $N_4$.
We have chosen to set\footnote{In the MC simulations we control the number of the $(4,1)$ simplices, see footnote \ref{foot1}, and the number of $(3,2)$ simplices adjusts dynamically.
For given values of $\kappa_0, \Delta$ the ratio of the two types of simplices is (approximately) constant and independent of $N_4$.}
$N_4^{(4,1)} =  160\,000$ and performed a series of measurements in a grid of points: $\kappa_0 = 3.0,...,4.5$, $\Delta= -0.07, ...,0.20$, see Figure~\ref{fig3}. In each point, we measured $\langle N_3 (t) \rangle$ and $\langle N_3(t_1) N_3(t_2)\rangle$ and used these data to calculate $\omega(\kappa_0,\Delta)$ and $\Gamma(\kappa_0,\Delta)$. The values of $\omega$ were obtained by fitting $\langle N_3 (t) \rangle$ to the semi-classical solution for an (elongated) four-sphere, see Eqs. \rf{j14}-\rf{j12}.  As in the volume profile $\langle N_3 (t) \rangle$ one observes a  "stalk" part, where $\langle N_3(t) \rangle \approx \mathrm{const.} $ in some range of $t$, and the "blob" part, where the semiclassical solution is valid, we  fitted a function\footnote{The existence of the "stalk" connecting both sides of the "blob" is due to the topological constraints of triangulations used in the MC simulations. Due to large volume fluctuations close to the $A-C_{\rm dS}$ phase transition the volume $\sum_t \langle N_3(t)\rangle $ contained in the "stalk"  is large making the volume $\sum_t \langle N_3(t)\rangle $ contained in the "blob" much smaller than total volume $N_4^{(4,1)}$.} 
\beql{Eqfitomega}
\la N_3(t) \ra = \max\left(\mathrm{const.}, \frac{3}{4\om } N_4^{3/4}\, \cos^3 \left( \frac{ t}{\om N_4^{1/4}}\right)\right)
\eeq
where $\mathrm{const.}$ and $N_4$ are free parameters and we checked that fitted values of $N_4$ agree with $\sum_t \langle N_3(t) \rangle$ inside the "blob".
The values of $\Gamma$ were computed twofold:
\begin{enumerate}
\item[(1)] by reconstructing the effective action \rf{j15} from the (connected) correlator  $ \langle N_3(t_1) N_3(t_2)\rangle - \langle N_3(t_1) \rangle \langle N_3(t_2) \rangle $ (for details  we refer to \cite{AMBJORN2011144,HandbookJGS}), and
\item[(2)] directly from the amplitude of fluctuations (note that the form of the effective action \rf{j15} implies Gaussian three-volume fluctuations with amplitude $\delta N_3(t) \propto  \sqrt{\Gamma N_4} $).
\end{enumerate}
In Fig. \ref{fig2} we compare both methods of measuring $\Gamma$ for fixed $\Delta = 0$
and a choice of $\kappa_0$ using four different lattice volumes $N_4^{(4,1)}=80\,000, 160\,000, 480\,000, 720\,000$,
note that the  proportionality constant in method (2) is the same in all plots.
It is seen that both methods agree except very close to the phase boundary,
where for larger volumes the results of method (1) are visibly higher than those of the method (2).
In further analysis, we decided to take the average $\Gamma$ from (1) and (2).

The dependence of $\omega$, $\Gamma$ and $\Gamma\omega^2$ on $\kappa_0,\Delta$ for fixed $N_4^{(4,1)} =  160\,000$   is shown in Figure \ref{fig3}. It is seen that both $\omega$ and $\Gamma$ change only slightly as a function of $\Delta$, whereas they strongly depend on $\kappa_0$ ($\omega$ decreases with $\kappa_0$ while $\Gamma$ increases with $\kappa_0$).  In the study of the ultraviolet limit we are interested in a product $\omega^2 \Gamma(\kappa_0,\Delta, N_4)$ which should diverge as $\sqrt{N_4}$ for $N_4\to \infty$, see Eq. (\ref{jo1}),
and it seems that the only way to achieve this is by approaching the $A-C_{\rm dS}$ phase transition line,
where indeed $\omega^2 \Gamma$ substantially grows.
The non-trivial problem remains how one should move in the $\kappa_0,\Delta$ coupling constant space when approaching the phase transition. Since, as seen in Fig.~\ref{fig3} (bottom), $\omega^2 \Gamma$ is almost independent of $\Delta$ and grows with $\kappa_0$, we have chosen the simplest possible way by fixing $\Delta$ and only changing $\kappa_0\to \kappa_0^{\UV}(N_4)$. In the following, we provide evidence that this conjecture may indeed be correct.

\begin{figure}
    \centering
    \includegraphics[width=0.45\linewidth]{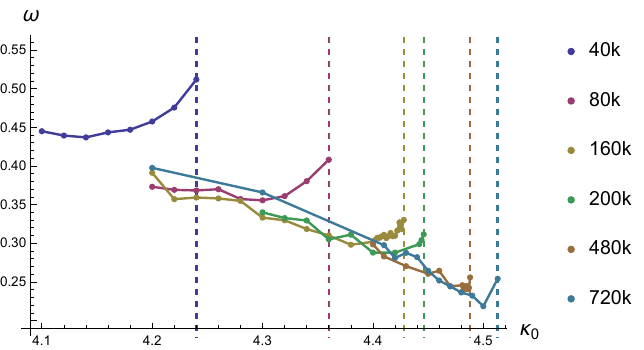}
    \includegraphics[width=0.45\linewidth]{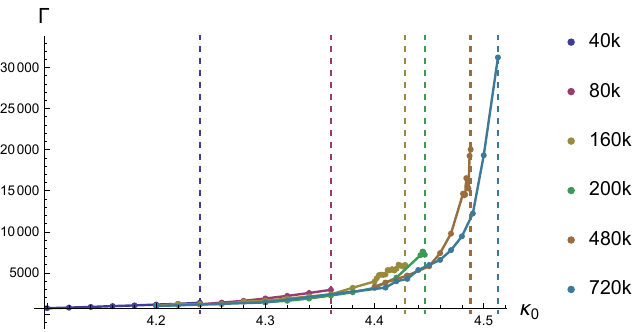}
    \includegraphics[width=0.45\linewidth]{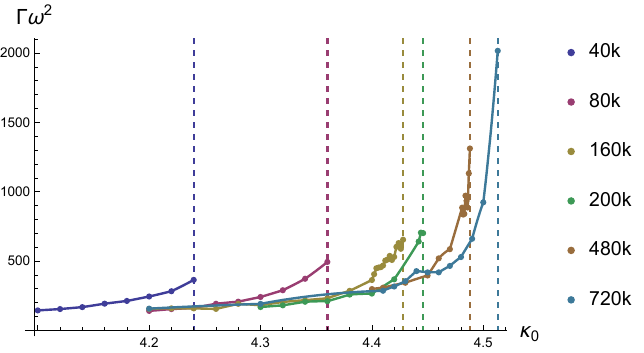}
    \caption{Dependence of $\omega$ (left), $\Gamma$ (right) and $\Gamma \omega^2$ (bottom) on $\kappa_0$ for fixed $\Delta=0$ measured for different lattice volumes $N_4^{(4,1)} = 40\, 000, 80\, 000, 160\, 000, 200\, 000, 480\, 000, 720\, 000$ (denoted by different colors). Positions of $\kappa_0$ closest to the pseudo-critical points $\kappa_0^{\UV}(N_4)$ are denoted by dashed lines.}
    \label{fig4}
\end{figure}

\begin{figure}
    \centering
    \includegraphics[width=0.45\linewidth]{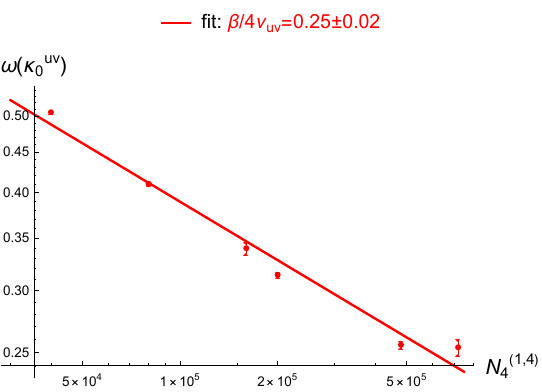}
    \includegraphics[width=0.45\linewidth]{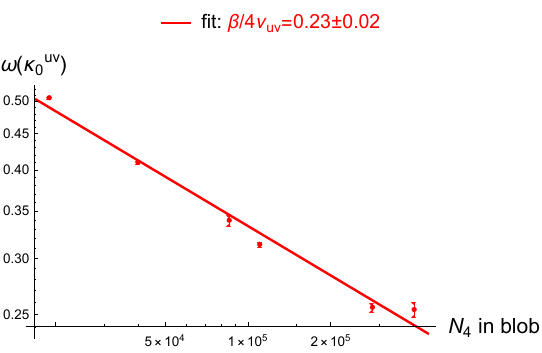}
    \includegraphics[width=0.45\linewidth]{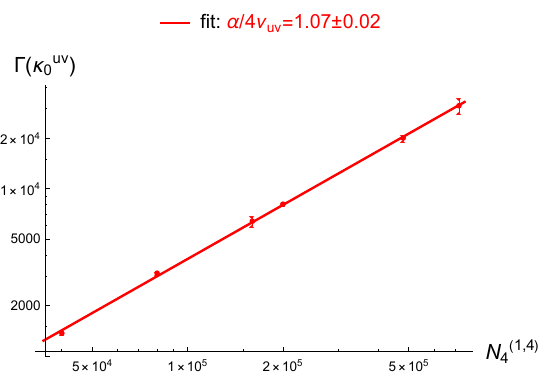}
    \includegraphics[width=0.45\linewidth]{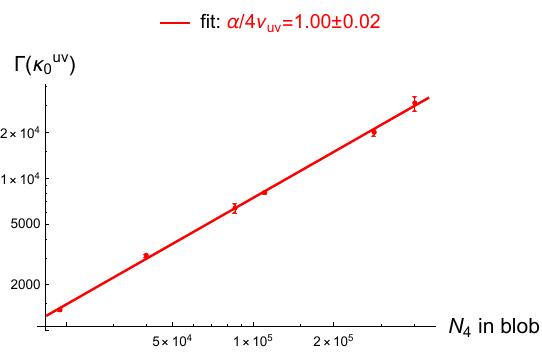}
    \includegraphics[width=0.45\linewidth]{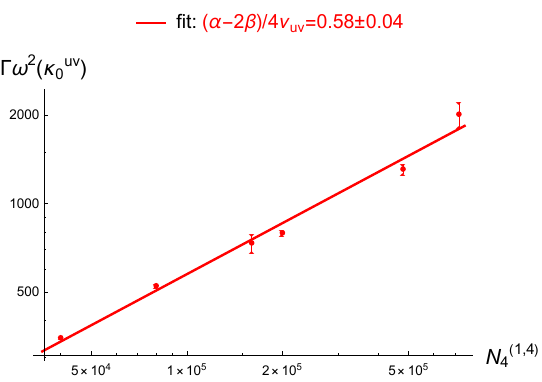}
    \includegraphics[width=0.45\linewidth]{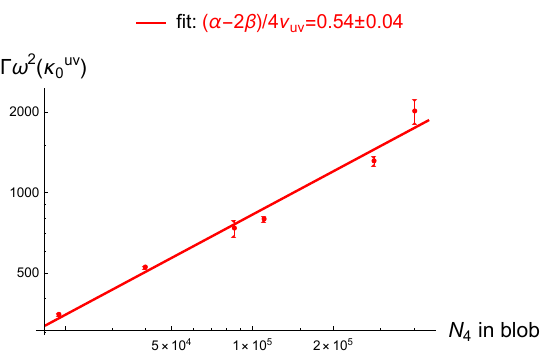}
    \caption{Critical scaling of $\omega$ (top), $\Gamma$ (middle) and $\Gamma \omega^2$ (bottom) measured closest to the pseudo-critical points $\kappa_0^{\UV}(N_4)$ (see Fig.~\ref{fig4}) for fixed $\Delta=0$. Fits of Eqs. \rf{jo4}-\rf{jo5} are depicted by solid lines. Left figures show scaling in the function of the total lattice volume $N_4^{(4,1)}$, right figures show scaling in the function of the $N_4$ volume contained in the "blob". Note the log-log scale in all plots.}
    \label{fig5}
\end{figure}

\begin{figure}
%\vspace{-2cm}
\centerline{\scalebox{0.9}{\rotatebox{0}{\includegraphics{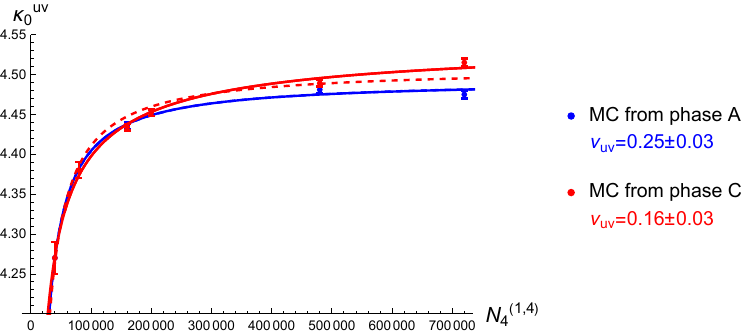}}}}
%\vspace{-5cm}
\caption{Critical scaling of $\kappa_0^{\UV}(N_4)$ at the $A-C_{\rm dS}$ phase transition. MC simulations initiated from inside phase $C$ are plotted in red and those initiated from inside phase are $A$ in blue. For the two largest volumes hysteresis is clearly visible. Solid lines are fits of Eq. \rf{jo3} including the critical exponent $\nu_{\UV}$ as a free parameter. For comparison we also plot as dashed lines the fits with fixed $\nu_{\UV} = 0.25$ which should be expected at a first order phase transition.}
\label{fig6}
\end{figure}

To investigate this in detail, we performed a series of precise MC measurements for fixed $\Delta = 0$ and a dense grid of $\kappa_0$ using a set of lattice volumes $N_4^{(4,1)} = 40\,000, 80\,000, 160\,000, 200\,000, 480\,000, 720\,000$.
In Figure \ref{fig4} we plot the measured values of $\omega$, $\Gamma$ and $\omega^2 \Gamma$.
It is seen that indeed well inside phase $C_{\rm dS}$ the behavior is independent of $N_4$, as it should be if the semiclassical solution \rf{j14}-\rf{j12} and the effective action \rf{j13} hold,
but very close to the phase boundary the measured values of $\omega^2\Gamma$ diverge more and more with increasing $N_4$.
Note that, as discussed in previous sections and in  Appendix 2, the position of pseudo-critical points also moves with $N_4$  and only in the limit $N_4\to\infty$ one recovers the true critical behavior when $\kappa_0^{\UV} (N_4) \to \kappa_0^{\UV}$.  One  can therefore  measure the dependence of $\omega$ and $\Gamma$ on $N_4$ at (or in practice as close as one can get to\footnote{We have chosen the largest $\kappa_0$ inside phase $C_{\rm dS}$ for which in the MC simulations we cannot see any tunneling between phase $C_{\rm dS}$ and phase $A$.})  pseudo-critical points  $\kappa_0^{\UV}(N_4)$ and  fit the critical exponents $\alpha, \beta$ and $\nu_{\UV}$ from Eqs. \rf{jo3}-\rf{jo5}. The results were summarized in Figures~\ref{fig5} and~\ref{fig6}. In Fig.~\ref{fig5} we show the critical scaling 
of $\om$, $\Gamma$ and $\om^2 \Gamma$ both as function of total lattice volume $N_4^{(4,1)}$ (left plots) and as  function of $N_4$ volume contained in the "blob" ($N_4$ values fitted from Eq. \rf{Eqfitomega}). The measured critical exponents from scaling with $N_4^{(4,1)}$ 
\beq\label{ceN4}
\frac{\beta}{4\nu_{\UV}} = 0.25 \pm 0.02, \quad \frac{\alpha}{4 \nu_{\UV}} 
= 1.07 \pm 0.02, \quad  \frac{\alpha-2\beta}{4 \nu_{\UV}} = 0.58\pm 0.04 
\eeq
are a bit higher than those resulting from scaling with $N_4$ in the "blob" 
\beq\label{ceN4blob}
\frac{\beta}{4\nu_{\UV}} = 0.23 \pm 0.02, \quad \frac{\alpha}{4 \nu_{\UV}} 
= 1.00 \pm 0.02, \quad  \frac{\alpha-2\beta}{4 \nu_{\UV}} = 0.54\pm 0.04 
\eeq
but agree within (two) standard deviations of the fits. In Fig.~\ref{fig6} we plot the critical scaling of $\kappa_0^{\UV}(N_4)$ at the $A-C_{\rm dS}$ phase transition together with a fit of Eq.~\rf{jo3}. The best fit of the critical exponent gives
\beq\label{cenuC}
\nu_{\UV} = 0.16\pm0.03.
\eeq
One should take this result with some caution as our earlier measurements showed that the $A-C_{\rm dS}$ transition is a first order phase transition \cite{cdt-finitesize,Ambjorn:2019pkp} for which one may expect $\nu_{\UV} = 0.25$. We have verified that for $\Delta=0$ one still observes many signatures of the first order transition such as MC simulation tunneling between two metastable states of phase $C_{\rm dS}$ and phase $A$ at pseudo-critical points. Also some hysteresis at the  position of the pseudo-critical points is observed for the two  largest  volumes in our MC simulations, see Fig.~\ref{fig6} where we compare  simulations initiated from inside phase $C_{\rm dS}$ (used above) with those initiated from inside phase $A$, the later data show the critical exponent
\beq\label{cenuA}
\nu_{\UV} = 0.25\pm0.03
\eeq
fully consistent with a first order transition.

\begin{figure}
    \centering
    \includegraphics[width=0.45\linewidth]{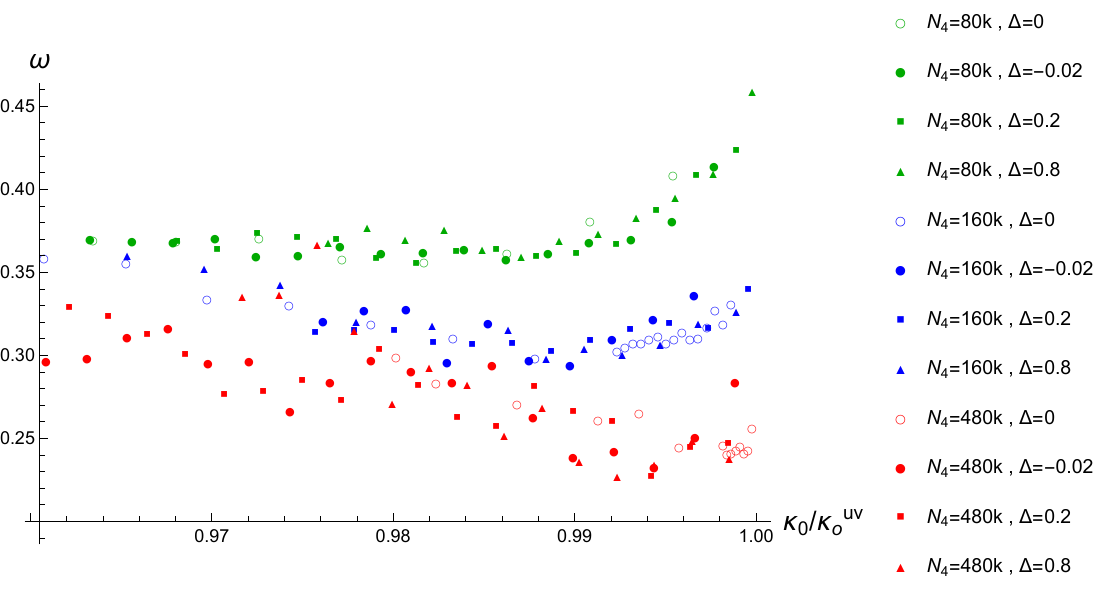}
    \includegraphics[width=0.45\linewidth]{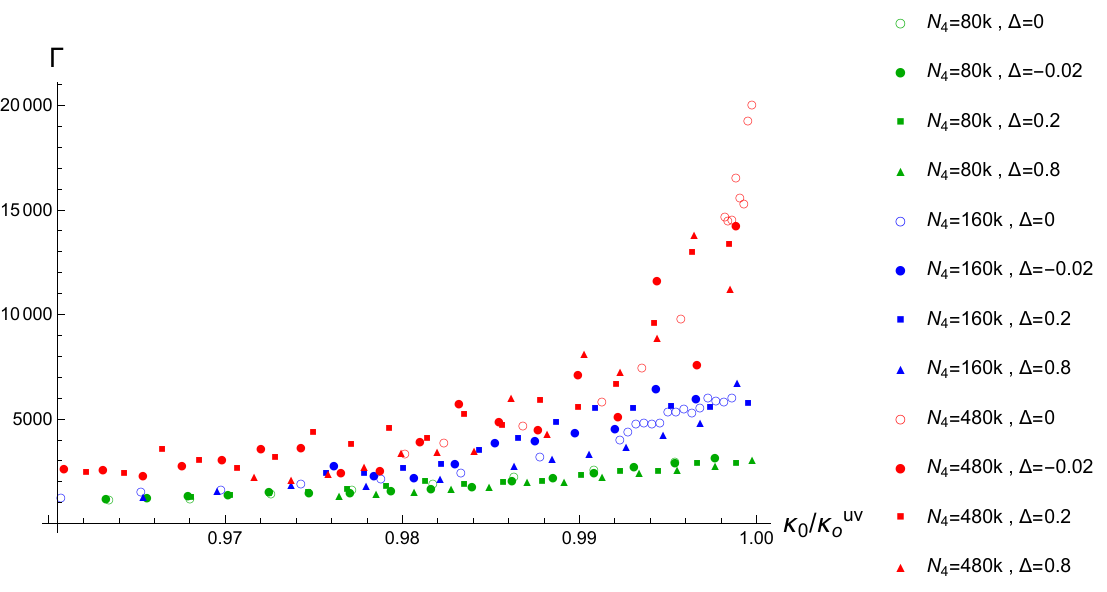}
    \caption{Universal behavior of $\omega$ (left) and $\Gamma$ (right) as a function of the (reduced) $\kappa_0/  \kappa_0^{\UV}(\Delta, N_4)$. Data measured for $\Delta=-0.02, 0, 0.2, 0.8$ are plotted with different markers. Colors indicate lattice volumes $N_4^{(4,1)} =  80\, 000$ (green), $160\, 000$ (blue) and  $480\, 000$ (red).}
    \label{fig:omega_gamma_unversal}
\end{figure}

The finite-size scaling analysis above was based on the assumption that we keep $\Delta$ fixed ($\Delta=0$) while changing $\kappa_0 \to \kappa_0^{\UV}(N_4)$. Last but not least, one should therefore investigate the dependence of the results on the choice of $\Delta$. We performed a series of similar measurements for $\Delta = -0.02, 0.2, 0.8 $. Due to the limited computer resources in these MC simulations, we only managed to investigate a smaller set of lattice volumes $N_4^{(4,1)}=80\, 000, 160\, 000, 480\, 000$. In Figure \ref{fig:omega_gamma_unversal} we plot $\omega$ and $\Gamma$ measured for various $\Delta$ as a function of the (reduced) $\kappa_0 / \kappa_0^{\UV}(\Delta, N_4)$, note that the position of pseudo-critical points $\kappa_0^{\UV}(\Delta, N_4)$ depends both on $\Delta$ and $N_4$.  Our conclusion is that the results are universal, independent of  $\Delta$ choice\footnote{These MC runs lasted much shorter than for $\Delta=0$ case. The smaller systems with $N_4^{(4,1)}=80\,000, 160\,000$ show very consistent results, while the data quality for the largest systems with $N_4^{(4,1)}=480\,000$ is visibly worse.}.

\section{Discussion}

In this article, we have tried to relate the simplest FRG flow to the CDT effective action for the scale factor.
Keeping the dimensionless lattice couplings fixed and moving towards the critical surface of the lattice theory,
the renormalized coupling constants will generally flow to an infrared fixed point.
Quite encouraging, we have seen a behavior that can be interpreted that way for the coupling constant $g_k\lam_k =G_k \Lam_k$,
where $G_k$ and $\Lam_k$ are the gravitational and the cosmological couplings used in FRG.
This infrared fixed point could either be the Gaussian fixed point observed in all FRG approaches,
or it could be the new IR fixed point discovered in \cite{frank-w}.
In principle, the choice of bare CDT coupling constants could lead either to the Gaussian fixed point or to the new IR fixed point,
such that we have a situation like the one illustrated in Fig.~\ref{fig-appendix} in  Appendix 2,
where there are two IR fixed points and one UV fixed point.
In the FRG approach, the flow to the Gaussian fixed point is very special and the generic flow would be a flow to the IR fixed point.
The present study cannot distinguish between the two fixed points.
However, it should be possible to do so, since the Gaussian fixed point allows for the value of the running gravitational coupling constant $G_k$ to be different from zero for $k \to 0$.
Approaching the Gaussian fixed point, one can then have a non-trivial coupling to matter fields.
That seems more difficult when approaching the IR fixed point where $G_0 = 0$.
In the past, we have studied the inclusion of scalar fields in four-dimensional CDT (and also in 2d CDT \cite{Ambjorn:2012kda}).
In four dimensions the effect has not been large, except when there has been a non-trivial interplay between the topology of spacetime
and topological properties of the scalar field \cite{Ambjorn:2021fkp}.
The large $N_4$ limits of these earlier investigations have to be reassessed in order to relate them to the flow towards an IR limit.

The analysis of the RG flow to a lattice UV fixed point provided in  Appendix 2 was based on the assumption
that one knows observables that have a well defined continuum limits when one approaches the UV fixed point.
In the FRG framework, such a dimensionless variable is $\Lam G$.
The FRG study, using the simplest truncation of the quantum effective action,
suggests that we can relate the running renormalized coupling constant $\Lam_k G_k$ to a similar quantity measured on the CDT lattice via Eq.~\rf{jh2a}.
This led us to identify possible UV fixed points in the CDT $\kp_0, \Del$ coupling constant space as located on the $A \mi C_{\rm dS}$ line.
We move along the transition line mainly by increasing the coupling constant $\Del$ (and making minor adjustments of $\kp_0$),
but the critical behavior seems to be quite independent of $\Del$ along the transition line.
This is why we dropped any reference to $\Del$ when discussing the approach to the $A \mi C_{\rm dS}$ line in the former Section.
Keeping $\Del$ fixed, we will approach the critical point $\kp_0^{\UV}(\Del)$ of the $A \mi C_{\rm dS}$ line
when following the pseudo-critical point $ \kp_0^{\UV}(\Del,N_4)$, taking $N_4 \to \infty$.
It is somewhat remarkable that following this pseudo-critical line $\om^2 \G$ scales approximately like $\sqrt{N_4}$, since $\om$ and $\G$ both scale quite differently and non-trivially when following $\kp_0^{\UV}(\Del, N_4)$.
This would also imply that $\kp_0^{\UV}(\Del)$ is indeed a candidate for a UV fixed point.
However, the present accuracy of our data does not allow us to claim with certainty that $\om^2 \G$ scales this way, and as discussed in the last Section, such a scaling is not needed in order to view $\kp_0^{\UV}(\Del)$ as a UV fixed point. What is needed is that $\om^2 \G$ scales with a power $N_4^\gamma$ where $\gamma \geq \oh$. If $\gamma > \oh$ one will be able to verify this with sufficient computer resources. Alternatively, if we eventually will find that $\gamma < \oh$ along the pseudo-critical line we would have to conclude that $\kp^{\UV}_0$ is not a UV fixed point and we would have the $\phi^4$ situation discussed in Appendix 2: we have a critical surface, but it is only associated with an IR fixed point.

The last point we would like to emphasize is the interpretation of $N_4^{1/4}$ as a correlation length that diverges as $N_4 \to \infty$. The numerical expressions for $\la N_3(t_i) \ra_{N_4}$ and $\la N_3(t_i) N_3(t_j)\ra_{N_4}$ fit perfectly with a finite-size scaling picture where one has a the correlation length $\xi$ equal to the size of the system, here being $N_4^{1/4}$. As discussed above, the correlation length refers to a point-point correlation. Such a correlation is only non-trivial and interesting in the case where we have fluctuating geometries, i.e. in quantum gravity, and in quantum gravity it is in some sense the most elementary diffeomorphism invariant correlator one can have. It contains a wealth of information about the quantum universe, e.g.\ about the local and global Hausdorff dimensions of the ensemble of geometries that enters into the path integral of quantum gravity. Since this correlator  is not generally appreciated let us for completeness provide the continuum definition here\footnote{The definition can be generalized to correlators not between points $x$ and $y$, but between fields $\phi(x)$ and $\phi(y)$ by including the matter action and also integration over field configurations in the path integral and replacing $1(x) 1(y)$ in \rf{dis1} with $\phi(x) \phi(y)$. It would then be the correlator $\la \phi(x) \phi(y)\ra$ where $x$ and $y$ are separated a geodesic distance $R$, averaged over all points $x,y$.}:
\beq\label{dis1}
G(R) =\!\! \int \cD [g]\; \e^{-S[g]}\! \int\!\!\!\int  d^4x d^4y \sqrt{g(x)} \sqrt{g(y)} 1(x) 1(y)\; \del (D_g (x,y)\mi R),
\eeq
where $D_g(x,y)$ denotes the geodesic distance between $x$ and $y$, calculated using the metric $g$,  and where $1(x)$ is the function with value 1 for all $x$. In the lattice formulation of quantum gravity we have in the path integral an ensemble of geometries. The (fractal) properties of this ensemble is determined by a combination of the Boltzmann weight associated to  each geometry via the action, and the entropy of the  geometries, i.e.\ the number of lattice geometries with the same action.  Depending on the choice of the bare coupling constants in the (lattice) action the ensemble of geometries might  or might not have interesting (fractal) properties. In the former case $G(R)$ will typically have a correlation length $\xi \propto \la N_4 \ra^{1/d_h}$, $d_h$ being the Hausdorff dimension of the ensemble of geometries. Thus $\xi \to \infty$ when $\la N_4 \ra \to \infty$. As we have mentioned above, this is a somewhat  different kind of criticality than one usually encounters in lattice field theory, where $\xi$ and $N_4$  will be independent. It is therefore not necessarily related to whether or not a phase transition (like the $A\mi C_{\rm dS}$) transition is a first or second order transition when viewed in terms of some other order parameters. This is why we do not discard the $A\mi C_{\rm dS}$ transition as relevant for continuum gravitation physics. Also, depending on the lattice gravity model, one can have such a scaling and a ``continuum limit'' that however has nothing to do with our ``real world''. As an example, consider  the so-called Dynamical Triangulations (DT) model where one is not imposing the time foliation present in CDT \cite{aj1,aj2}. The model has two coupling constants $\kp_0$ and $\kp_4$, corresponding to the similar coupling constants in CDT. There is a phase transition point $\kp_0^c$ in $\kp_0$ and for $\kp_0 > \kp_0^c$ one finds a so-called branched polymer phase: for a given value of $\kp_0 > \kp_0^c$ the statistical model becomes critical for $\kp_4 \to \kp_4^c(\kp_0)$. One finds that $\la N_4 \ra \to \infty$ as $1/(\kp_4 -\kp_4^c(\kp_0))$ and a correlation length $\xi(\kp_4) \propto \la N_4 \ra^{1/2}$, determined by the exponential fall off of $G(R)$. We have a universe consisting of branched polymers with Hausdorff dimension 2, and one can take a scaling limit but it is not interesting from the point of view of representing our 4d world\footnote{One can try to analyze the DT model by approaching the critical point $\kp_0^c$ either from the branched polymer phase or from the other side, the so-called crumpled phase and try to find non-trivial scaling, precisely like we here find non-trivial scaling when approaching the $A\mi C_{\rm dS}$ transition from the $C_{\rm dS}$ side. This is still work in progress and also includes a study of models that enlarge the pure Einstein-Hilbert action by adding other terms (somewhat similar to the additional terms in the  CDT action) \cite{jack,jack1}}. The remarkable aspect of the CDT geometries in phase $C_{\rm dS}$ is not that they exhibit finite-size scaling (this is also the case for the DT branched polymers), but that they  look like a de Sitter universe with superimposed quantum fluctuations, and the exciting aspect of the present results is that there is a reasonable chance that the lattice theory, apart from a IR limit, also has a non-trivial UV limit, but clearly the quality of the data is such that this is not settled yet, and it is fair ending the article answering the question mark in the title with a ''maybe''.

\subsection*{Acknowledgments}
It is a pleasure to thank Frank Saueressig for numerous enlightning discussions about FRG. We are also grateful to Renata Ferrero and Martin Reuter for discussing the physics related to the simplest FRG truncation used in this article as well as Kevin Falls and Renata Ferrero for discussions related to the critical UV exponent. We also thanks Renate Loll for a careful reading of the manuscript and corresponding constructive comments. JA thanks the Perimeter Institute for Theoretical Physics, where part of this work was completed, for hospitality and support. Research at the Perimeter Institute is supported by the Government of Canada through the Department of Innovation, Science and Economic Development and by the Province of Ontario through the Ministry of Colleges and Universities. DN is supported by the VIDI programme with project number VI.Vidi.193.048, which is financed by the Dutch Research Council (NWO). The research has been supported by a grant from the Priority Research Area (DigiWorld) under the Strategic Programme Excellence Initiative at Jagiellonian University. 

\section*{Appendix 1}

In this Appendix we summarize for completeness the notation used in CDT and 
compare it with the continuum notation we use for FRG. Below $a$ denotes 
the length of the lattice links. In CDT we work with triangulations made from gluing together four-simplices. The triangulation has $N_t$ time slices, and each time slice $i$ consists of $N_3(i)$ tetrahedra glued together to form a triangulation with the topology of $S^3$. The number of four-simplices filling the slab between time slice $i$ and time slice $i\! +\!1$ is denoted $N_4(i)$. This setup should be compared with a continuum hyperbolic spacetime using proper-time coordinates where the spatial topology is that of $S^3$ and where the spatial volume is $V_3(t)$
\beq\label{appp1}
N_3(i) a^3 \propto V_3(i) \to V_3(t_i), \quad t_i = i \, a ,
\eeq
\beq\label{appp2}
N_4 a^4 = \sum_{i=1}^{N_t} \;N_4(i)a^4 \to \int^{t_{N_t}}_{t_{1}} dt \;V_3(t) = V_4.
\eeq
The volume of a tetrahedron with side-lengths 1 is $\sqrt{2}/12$, while the volume of a four-simples is $\sqrt{5}/96$, but in the article we set these factors to 1 for notational  simplicity. The factors will play no role in the scaling arguments. In the slab between $i$ and $i+1$ we have two kinds of four-simplices, the $(3,2)$ simplices, and the $(4,1)$ simplices.  We denote the total number of these: $N_4^{(3,2)}$ and  $N_4^{(4,1)}$, respectively. Each of the $N_3$ tetrahedra at the time-slices is connected to exactly two $(4,1)$ simplices. %Only the $(4,1)$ simplices are connected to the $N_3$ tetrahedra at the time-slices. 
Thus we have 
\beq\label{appp3}
2 \sum_{i=1}^{N_t} N_3(i) = N_4^{(4,1)}  \to  c\, N_4 \quad {\rm for } \quad N_4 \to \infty.
\eeq
The constant $c$ depends on the coupling constants $\kappa_0,\Delta$ in phase $C_{\rm dS}$, reflecting the ratio between $N_4^{(3,2)}$ and $N_4^{(4,1)}$. Again it will play no role in the scaling arguments, and we have again just put it to 1 for notational simplicity. Thus we write, a little sloppily:
\beq\label{appp4}
\sum_{i=1}^{N_t}  a \, N_3(i)a^3  = N_4 a^4 \to \int^{t_{N_t}}_{t_{1}}  dt \;V_3(t) = V_4.
\eeq
The Regge action for a triangulation $T$ with $N_4(T)$ building blocks becomes amazingly simple, depending only on the global number of  simplices and sub-simplices in the triangulation. Let $N_2(T)$ denote the number of triangles and $N_0(T)$ the number of vertices in $T$. Then the Regge action will be 
\beq\label{appp6}
 S[T] = -\tilde{\kappa}_2 N_2(T) + \tilde{\kappa}_4 N_4(T) = 
 -\kappa_0 N_0(T) + \kappa_4 N_4(T),
 \eeq
where the last equality sign is correct up to a term involving the Euler constant for the four-manifold. This term plays no role for large $N_4$.

This formula {\it is} the Einstein-Hilbert action as introduced by Regge, and the relation to the coupling constants $G$ and $\Lambda$ is 
 \beq\label{appp7}
 \kappa_0 = c_1 \frac{a^2}{G}, \qquad \kappa_4 = c_2 \frac{a^2}{G} + 
 c_3 \frac{\Lambda}{G} \, a^4.
 \eeq
 The precise constants can be found in \cite{review1}. In CDT we generalize this action a little, in the Wilsonian spirit,  by attributing different cosmological constants to $N_4^{(3,2)}$ and $N_4^{(4,1)}$. Thus we write
 \beq\label{appp8}
 S[T] = -\kappa_0 N_0(T) + \kappa_{32} N_4^{(3,2)}(T) + \kappa_{41} N_4^{(4,1)}(T).
 \eeq
This can also be written in the following parametrization
\beq\label{appp9}
 S[T] = -(\kappa_0 +6 \Delta) N_0(T)  + \kappa_4 N_4(T) + \Delta N_4^{(4,1)}(T),
\eeq
where $\Delta=0$ corresponds to assigning the same cosmological constant to $(3,2)$ and $(4,1)$ simplices. This is the origin of the bare coupling constant space $\kappa_0,\Delta$ on the lattice and their relation to the continuum coupling constants $G$ and $\lambda$. 
The important relation is $\kappa_0 \propto a^2/G$.

\section*{Appendix 2}

\subsubsection*{The {$\pmb{\phi^4}$} example}
\begin{figure}[t]
\centerline{\scalebox{0.2}{\rotatebox{0}{\includegraphics{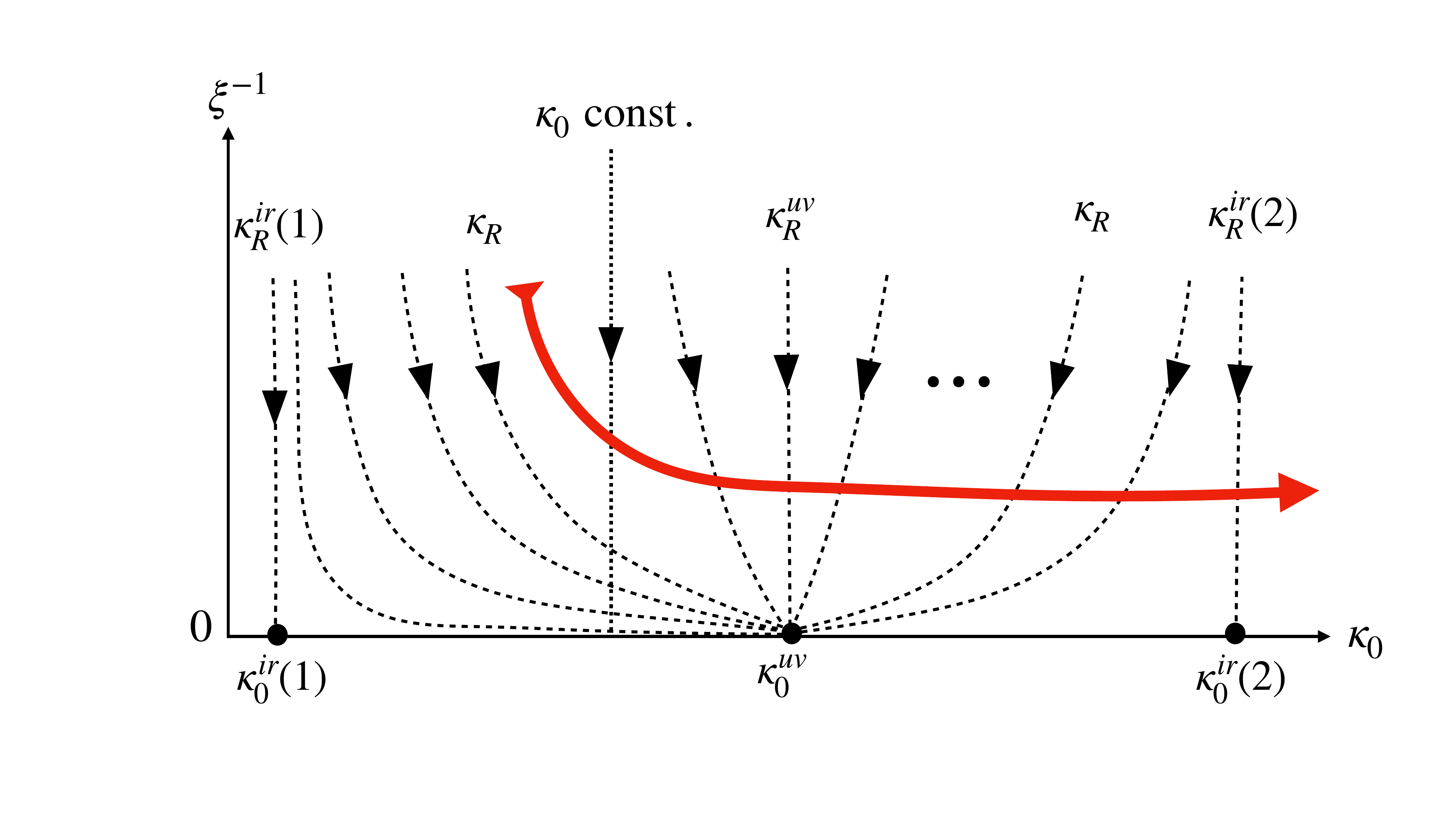}}}}
\caption{\small The tentative $\phi^4$ phase diagram with 
an UV fixed point and two IR fixed points. The dashed lines are paths where the renormalized $\phi^4$ coupling constant $\kp_R$ is kept fixed, while on the dotted line  the bare coupling constant $\kp_0$ is fixed. The fulldrawn red line illustrate the 
way the real renormalization group flow would be in a $\phi^4$ theory with a fixed $\kp_R$. It would never reach the critical line where $\xi = \infty$, and accordingly there would not be a continuum quantum field theory with a fixed $\kp_R > 0$.}
\label{fig-appendix}
\end{figure}

As an example of how to approach a lattice field theory UV fixed point,
we consider a $\phi^4$ lattice field theory in four dimensions.
\footnote{Disclaimer: the $\phi^4$ theory in four dimensions has no UV fixed point, so the scenario outlined is not realized,
as mentioned in the figure caption of Fig.\ \ref{fig-appendix}.
However, the fact that this scenario was not seen in a $\phi^4$ theory was used in \cite{luescher}
to argue that there is no UV fixed point in the $\phi^4$ lattice field theory.}
The lattice action used is 
\beq\label{f1}
S = \sum_n \left(  \sum_{m=1}^4 (\phi(n \plu m) \mi \phi(n))^2 + \mu_0 \phi^2(n) + \kp_0 \phi^4(n) \right),
\eeq
where the field and the coupling constants $\mu_0$ and $\kp_0$ are dimensionless and the length of the lattice links is $1$. We assume $\kp_0 \geq 0$.
At each value of the coupling constants $\mu_0, \kp_0$ we have a correlation length $\xi(\mu_0, \kp_0)$ of the two-point function.
The lattice theory has a second order phase transition line where $\xi = \infty$.
It is the phase transition line separating the symmetric phase where $\la \phi \ra = 0$ from the broken phase where $\la \phi \ra \neq 0$.
We will only discuss approaching the phase transition line from the symmetric phase.
Rather than using $\mu_0$, it is convenient to use $\xi^{-1}$ as a variable when discussing the phase diagram.
The phase transition line will then be the line $\xi^{-1} = 0$.
On this phase transition line there might be UV and IR fixed points, as illustrated in Fig.~\ref{fig-appendix}.
In a $\phi^4$ theory in four dimensions, the existence of a UV fixed point would imply that there exists a renormalized $\phi^4$ coupling constant $\kp_R > 0$. This renormalized coupling constant is usually defined in a specific way in terms of correlation functions at a given momentum scale and it is an observable in the continuum renormalized quantum field theory. All renormalized observables   $\mathcal{O}_R$ in the quantum field theory can be expressed in terms of the bare coupling constants and the cut-off used to define the theory. Here we use a lattice regulatization and the cut-off is the lattice spacing $a$, which we in \rf{f1} have put to 1. For all renormalized observables $\mathcal{O}_R ( \kp_0,a)$, the value should be independent of $a$ when $a$ is small (compared to any physical length scale in the theory). This will only be possible if we change $\kp_0$ at the same time as we change $a$. Thus we have to write $\mathcal{O}_R ( \kp_0(a),a)$, where 
\beq\label{fj2}
0 = a\frac{d }{d a} \mathcal{O}_R( \kp_0(a),a)=
\Big(a\frac{\prt}{\prt a } + a\frac{d\kp_0(a)}{da}
\frac{\prt}{\prt \kp_0} \Big) \mathcal{O}_R( \kp_0(a),a).
\eeq
This is the simplest Callen-Symanzik equation. 
In particluar we can apply it to $\kp_R(\kp_0(a),a)$.
Finally, we can relate the cut-off $a$ to the correlation
length $\xi$ by insisting that the physical correlation
length $\ell_{phys} = \xi a$ stays fixed when $a \to 0$.
This is equivalent to saying that the lattice exponential 
decay of the two-point function survives in 
the continuum when $a \to 0$, and that this exponential
decay, which defines the renormalized mass $m_R$ in the 
$\phi^4$ theory, is given by 
\beq\label{fj3}
\frac{1}{\xi} = m_R a. 
\eeq
Introducing $\xi$ instead of $a$ in the 
Callen-Symanzik equation \rf{fj2} we obtain
\beq\label{fj4}
0 = \xi\frac{d }{d \xi} \kp_R( \kp_0(\xi),\xi)=
\xi \frac{\prt \kp_R}{\prt \xi }\Big|_{\kp_0} + 
\frac{\prt \kp_R}{\prt \kp_0}\Big|_\xi\;
\xi\frac{d\kp_0}{d\xi}\Big|_{\kp_R}
\eeq
The curves $\kp_R( \kp_0(\xi),\xi) = \const$ are the ones
shown in Fig.~\ref{fig-appendix} for different choices of the value of $\kp_R$.
The renormalized value $\kp_R$ is fixed not at the UV fixed point $\kp_0^{\UV}$,
but by the approach to it and the same is true for the other renormalized 
coupling constant, $m_R$. In this sense $\kp_R$ and $m_R$ are 
free parameters in the renormalized theory.

Introducing the $\beta$-functions\footnote{Only when we are close to the critical line, i.e.\ when $\xi$ is very large, will $\xi \prt \kp_R (\kp_0(\xi), \xi)/\prt \xi$ be a function only of $\kp_R$. A similar statement is true for $\xi d\kp_0/d\xi$. We will assume we are in this so-called scaling region of coupling constant space. The reason for the different sign appearing in the two equations in \rf{fj5} is that the change in $\kp_R$ is usually defined wrt a change of a physical mass scale like $m_R$ in \rf{fj3}.}
\beq\label{fj5}
\beta_0(\kp_0) = \xi\frac{d\kp_0}{d\xi}\Big|_{\kp_R}, \qquad
\beta_R(\kp_R) =-\xi \frac{\prt \kp_R}{\prt \xi }\Big|_{\kp_0},
\eeq
Eq.~\rf{fj4} can be written as 
\beq\label{fj6}
\beta_R(\kp_R) = \frac{\prt \kp_R}{\prt \kp_0} \;\beta_0(\kp_0)
\eeq
and the two $\beta$-functions will have the same  sign, but the sign difference in the definition \rf{fj5} means that the behavior of $\kp_0(\xi)$ and $\kp_R(\xi)$ will be different when one solves Eq. \rf{fj5} for $\xi \to \infty$, i.e.\ when approaching the critical line. The $\beta$-function $\beta_0(\kp_0)$ in our $\phi^4$ example is assumed to be zero at the fixed points and positive in the interval $]\kp_0^{\IR}(1), \kp_0^{\UV}[$ (and negative in the interval $]\kp_0^{\UV},\kp_0^{\IR}(2)[$). Thus  $\kp_0 \to \kp_0^{\UV}$ when $\xi \to \infty$, (and by definition one follows a path where $\kp_R$ is kept fixed). On the other hand, solving \rf{fj5}
for $\kp_R$ we find that $\kp_R$ will be repelled by $\kp_R^{\UV}$ and move towards one of the IR fixed points when $\xi \to \infty$, as one can see in Fig.\ 
\ref{fig-appendix}. As a simple toy example to illustrate
the behavior let us assume the $\beta$-function for the 
$\phi^4$ is given by 
\beq\label{fj7}
\beta(\kp) = \frac{c_0 \kp^2 (\kp^{\UV}-\kp)}{\nu c_0 \kp^2 + (\kp^{\UV}-\kp)}, 
\quad \beta_0(\kp_0) = \beta(\kp_0), \quad \beta_R(\kp_R) =
\beta(\kp_R).
\eeq
This toy $\beta$-function imitates the real $\phi^4$ $\beta$-function to first order at the Gaussian IR fixed point $\kp_0^{\IR} = \kp_R^{\IR} = 0$ and has a UV fixed point at $k^{\UV}$. Also, it is a rational function of the coupling constant, a feature also present in the simplest FRG truncations.\footnote{In general $\beta_0(\kp_0)$ and $\beta_R(\kp_R)$ will not be identical, but one can show that  the two first coefficients when expanding around a Gaussian  fixed point are the same. Further, if the derivatives of the  $\beta$-functions are different from zero at the UV fixed point (like here) they have to agree.}  One can now solve for $\xi$ and find
\beq\label{fj8}
\xi(\kp_0,\kp_R) = \frac{f(\kp_R)}{f(\kp_0)}, \quad 
f(\kp) = \e^{1/c_0 \kp} (\kp^{\UV} -\kp)^\nu, \quad 
0 < \kp < \kp^{\UV}.
\eeq
For fixed $\kp_R$ it will reproduce a sequence of curves like the ones shown at the left side of $\kappa_0^{\UV}$ in Fig.\ \ref{fig-appendix}. Let us finally note that if the IR fixed point is characterized by an exponent $\nu_{\IR}$ and the UV fixed point by an exponent $\nu_{\UV}$ a toy model beta function 
will be 
\beq\label{fj7a}
\beta (\kp) = \frac{(\kp - \kp^{\IR})(\kp^{\UV}-\kp)}{\nu_{\IR} (\kp^{\UV}-\kp) + \nu_{\UV}(\kp - \kp^{\IR})},
\quad \beta_0(\kp_0) = \beta(\kp_0), \quad 
\beta_R(\kp_R) = \beta(\kp_R),
\eeq
and one finds
\beq\label{fj8a}
\xi(\kp_0,\kp_R) = \frac{f(\kp_R)}{f(\kp_0)},
\quad f(\kp) = 
\frac{(\kp^{\UV}-\kp )^{\nu_{\UV}}}{(\kp-\kp^{\IR})^{\nu_{\IR}}}, \quad \kp^{\IR} < \kp < \kp^{\UV}.
\eeq
It should be mentioned that the recently calculated
$\beta$-function for $\lam_k g_k$ in FRG has precisely 
the form \rf{fj7a}, except that it is more complicated
 far away from $\kp^{\IR}$ and $\kp^{\UV}$ \cite{kawai}.

Let us also mention that irrespectively of the detailed form of the $\beta$-function, if we on the lattice observe a divergent correlation length $\xi$ for $\kp_0 \to \kp_0^{\UV}$ of the form
\beq\label{add1}
\xi \propto \frac{1}{(\kp_0^{\UV} - \kp_0)^{\nu_{\UV}}}, \quad 
{\rm i.e.} \quad \kp_0(\xi) = \kp_0^{\UV} - 
\frac{c}{\xi^{1/{\nu_{\UV}}}}
\eeq
then \rf{fj5} leads to 
\beq\label{add2}
\beta(\kp_0) \approx \beta'(\kp_0^{\UV}) (\kp_0 - \kp_0^{\UV}) =
- \frac{1}{\nu_{\UV}} (\kp_0 - \kp_0^{\UV}), \quad
{\rm i.e.} \quad \nu_{\UV} = - \frac{1}{\beta'(\kp_0^{\UV})}.
\eeq

To summarize: keeping the renormalized coupling $\kp_R$ fixed and taking the lattice correlation length to infinity, the lattice coupling constant $\kp_0$ flows to the UV fixed point $\kp_0^{\UV}$, while keeping $\kp_0$ fixed while taking the lattice correlation length to infinity results in a flow of the renormalized coupling constant to an IR fixed point $\kp_R^{\IR}$. 

\subsubsection*{Application to CDT}

We now want to apply this philosophy to the CDT lattice theory and we assume that the FRG effective action is expressed in terms of renormalized quantities. We have three dimensionless coupling constants, $\kp_4$, $\kp_0$ and $\Del$. $\kp_4$ is the coupling constant multiplying $N_4$, the number of four-simplices, and as discussed in the main text there is a $\kp_4^c(\kp_0,\Del)$ such that $\la N_4 \ra \to \infty$ for $\kp_4 \to \kp_4^c(\kp_0,\Del)$. We will be using $N_4$ as a variable instead of $\kp_4$, performing MC simulations for various values of $N_4$.
\begin{figure}[t]
%\vspace{-2cm}
\centerline{\scalebox{0.2}{\rotatebox{0}{\includegraphics{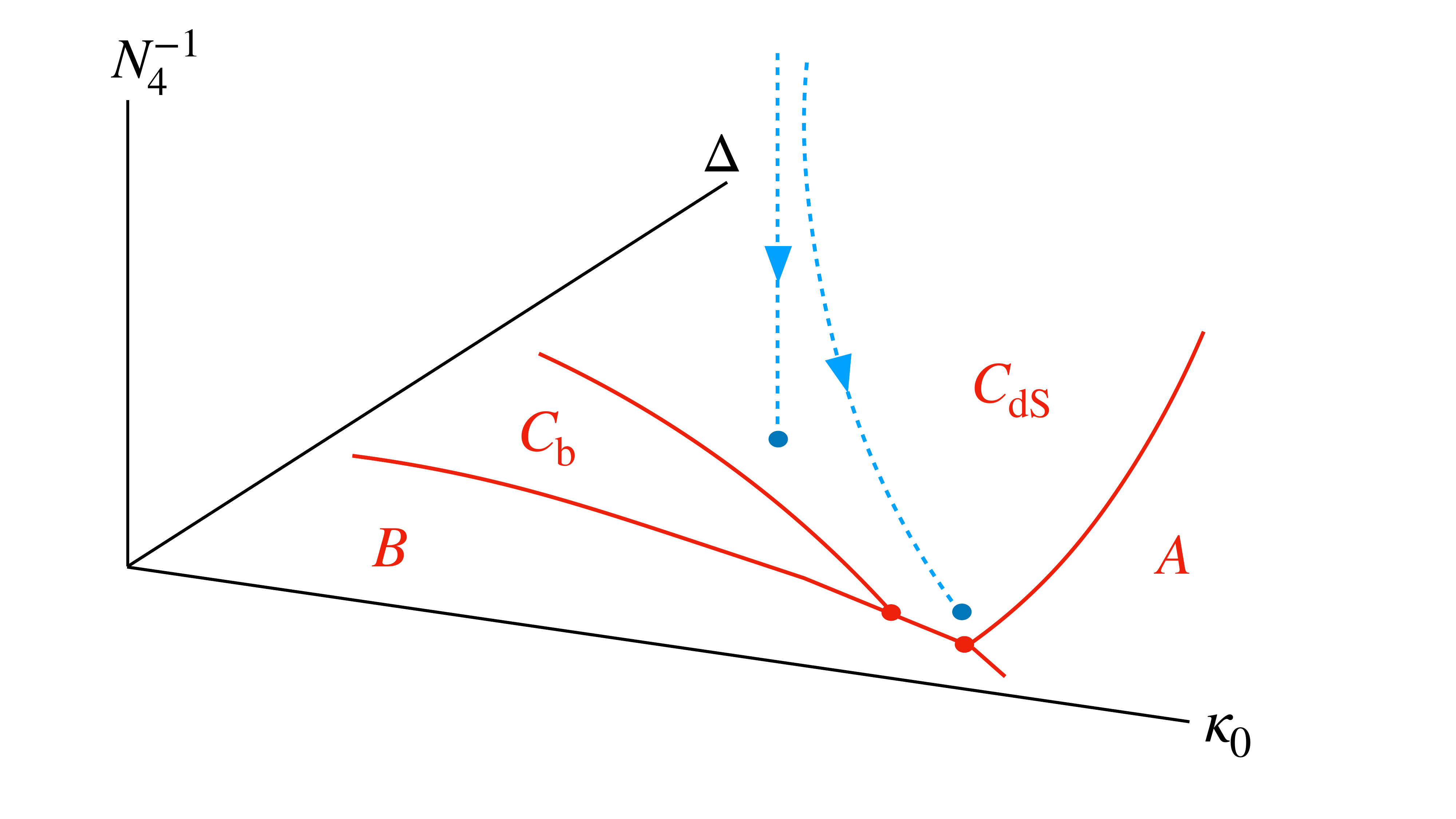}}}}
%\vspace{-5cm}
\caption{{\small The CDT phase diagram where $N_4^{-1}$ is 
also included. Criticality can only occur when 
$N_4^{-1}= 0$. The straight %horizontal 
vertical dashed line corresponds
to keeping the bare lattice coupling constants $\kp_0,\Del$ fixed, while the other dashed line  illustrates the flow when the renormalized coupling constants are fixed and one has to change the lattice coupling constants when approaching the critical surface.}}
\label{fig2-appendix}
\end{figure}
As discussed in the main text, the relevant correlation length related to the finite-size scaling observed in phase $C_{\rm dS}$
\cite{agjl1,cdt-finitesize}, is the point-point correlation, a correlation unique for quantum gravity and for fixed (and large) $N_4$ it will be proportional to a typical linear size of a universe of spacetime volume $N_4$, i.e.\
in the case of 4d CDT to $N_4^{1/4}$:
\beq\label{fj9}
\xi \propto N_4^{1/4}.
\eeq
The critical surface of the lattice theory will be for 
$N_4^{-1} \to 0$. The corresponding CDT phase diagram 
is shown in Fig.\ \ref{fig2-appendix} as a function of 
$\kp_0,\Del$ and $N_4^{-1}$. 
For fixed bare coupling 
constants $\kp_0,\Del$ we thus expect that the renormalized 
coupling constants will flow towards their values at an 
IR fixed point. Similarly, we expect that for fixed 
renormalized values of the coupling constants we have 
to change $\kp_0$ and $\Del$ as functions of $N_4$, when 
approaching the critical surface $N_4 = \infty$, in order 
to keep the renormalized coupling constants fixed. This 
is illustrated by the two curves shown in 
 Fig.\ \ref{fig2-appendix} that should be viewed as the 
 equivalent to the $\phi^4$ diagram shown in 
Fig.\ \ref{fig-appendix}. Unfortunately we do not know 
how to calculate the renormalized coupling constants
corresponding to the lattice theory, and as discussed in
the main text we just assume that they can be 
identified with the renormalized coupling constants 
in the simplest possible FRG model. Thus for fixed 
lattice coupling constants $\kp_0,\Del$ the basic equation
\rf{j16} tells us 
\beq\label{fj10}
\lam_k g_k \propto  \frac{\G}{\xi^2},
\qquad {\rm i.e.} \qquad 
\xi \frac{ d}{d \xi} (\lam_k g_k ) = -2 \lam_k g_k.
\eeq
This is indeed in agreement with \rf{js4} since near the Gaussian
fixed point or the IR fixed point we also have from \rf{j16}
that $\xi \propto k^{-2}$, more specifically (from 
\rf{ju} and \rf{ju4}):
\beq\label{fj11}
\xi \propto \frac{\sqrt{\G}}{G_0 k^2} \quad {\rm
(Gaussian)}, \qquad \quad\xi \propto \frac{\sqrt{\G} \,k_0^2}{k^2} \quad
{\rm (IR)}.
\eeq
In the notation of \rf{fj7a} and \rf{fj8a} we have an
IR fixed point with $\nu_{\IR} = 2$.

\begin{figure}[t]
%\vspace{-2cm}
\centerline{\scalebox{0.2}{\rotatebox{0}{\includegraphics{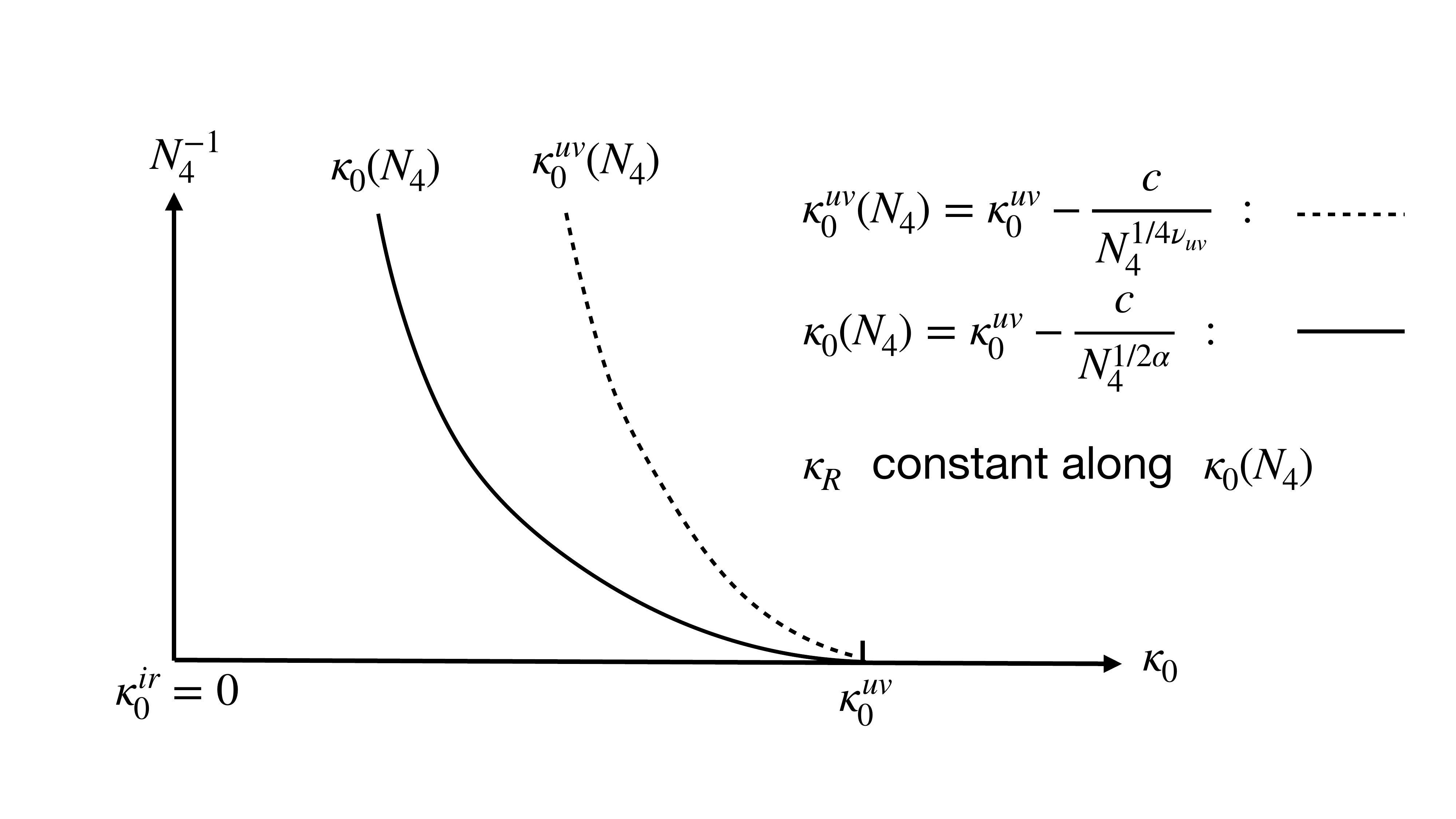}}}}
%\vspace{-5cm}
\caption{{\small The tentative CDT phase diagram $(\kp_0,N_4^{-1})$ (with coupling constant $\Del$ ignored). Pseudo-criticallity appears along the dotted line $\kp_0^{\UV}(N_4)$ and the line $\kp_0(N_4)$, where $\kp_R$ is constant is shown to the left of $\kp_0^{\UV}(N_4)$, $\kp_R \propto \lam_kg_k$. The critical line is $N_4^{-1} = 0.$}}
\label{fig3-appendix}
\end{figure}

Let us now discuss how to deal with the UV lattice fixed point. According to the discussion above we should now keep to renormalized coupling constant fixed, but it can be assigned any value between the IR value and the UV value. Viewing $\lam_kg_k$ for some value of $k$ as one of these values, Eq.~\rf{j16} implies (as already mentioned) that we should follow a path $(\kp_0(N_4),\Del(N_4))$ such that 
\beq\label{fj12}
\G(\kp_0(N_4),\Del(N_4), N_4) \propto \sqrt{N_4}\quad (\propto \xi^2).
\eeq
Let us for simplicity of the discussion  ignore the coupling constant $\Del$, and likewise we will for the same reason ignore that the observed $\om$ is not equal to $\om_0$ corresponding to a 4-sphere. This last point is discussed in detail in the main text. Finally let $\kp_0^{\UV}$ denote the putative UV fixed point with exponent $\nu_{\UV}$. Assume that $\G(\kp_0)$ has the following behavior on the critical surface 
$N_4 = \infty$, close to $\kp_0^{\UV}$:
\beq\label{fj14}
\G(\kp_0,N_4\equ \infty) \propto \frac{1}{(\kp_0^{\UV}-\kp_0)^{\a}} 
\eeq
Strictly speaking for a finite $N_4$ we do  not have a critical point $\kp_0^{\UV}$, but only a so-called pseudo-critical point $\kp^{\UV}_0(N_4) < \kp^{\UV}_0$ and we expect 
\beq\label{fj13}
\kp^{\UV}_0(N_4) = \kp_0^{\UV} - \frac{c}{N_4^{1/4\nu_{\UV}}}, 
\quad \xi \propto N_4^{1/4} \propto \frac{1}{(\kp_0^{\UV}-\kp_0(N_4))^{\nu_{\UV}}}. 
\eeq 
It is thus natural to identify the exponent $\nu_{\UV}$ introduced this way with the exponent $\nu_{\UV}$ we encountered in our $\phi^4$ discussion. Since for a given $N_4$ we cannot really choose $\kappa_0  > \kappa_0^{\UV}(N_4)$ there is one ``natural'' way to approach the UV fixed point $\kappa_0^{\UV}$ corresponding to $N_4 = \infty$, namely to follow the path $\kappa_0^{\UV}(N_4)$. In this case we obtain
\beq\label{fj14a}
\G(\kp_0^{\UV}(N_4),N_4) \propto \frac{1}{(\kp_0^{\UV}-\kp_0^{\UV}(N_4))^{\a}} \propto N_4^{\a/4\nu_{\UV}}.
\eeq
This will be an upper bound on $\G(\kp_0(N_4),N_4)$ since $\kp_0(N_4)$ is less than the pseudo-critical $\kp_0^{\UV}(N_4)$. Comparing this to \rf{fj12} we have 
to have 
\beq\label{fj20}
\frac{\a}{4\nu_{\UV}} \geq \frac{1}{2},
\eeq
and in order to approach the UV fixed point $\kappa_0^{\UV}$ such that \rf{fj12} is satisfied  one should choose a path 
\beq\label{fj21}
\kappa_0 (N_4) = \kappa_0^{\UV} - \frac{c}{N_4^{1/2\alpha}}.
\eeq
Eq.\ \rf{fj20} implies that we indeed have $\kappa_0 (N_4) \leq \kappa_0^{\UV}(N_4)$. The situation is illustrated in Fig.\ \ref{fig3-appendix}.

Assuming $\xi \propto N_4^{1/4}$ and using \rf{add2} makes it tempting to conjecture that the $\beta$-function for 
$\kp_0$ will satisfy 
\beq\label{add3}
\theta_{\kp_0^{\UV}} \equiv \beta'(\kp_0^{\UV}) = - \frac{2}{\alpha}
\eeq
We will use this in the main article, but it should be kept in mind that the behavior \rf{fj21} was obtained by assuming that the renormalized coupling $\Lam G$ was kept fixed, not the renormailzed version of the lattice coupling $\kp_0$, which we strictly speaking do not know. However, since $\theta_{\kp_0^{\UV}} = \theta_{f(\kp_0^{\UV})}$ for any reasonable function $f(\kp_0)$ where $f'(\kp_0^{\UV}) \neq 0$, the critical exponent $\theta_{\kp_0^{\UV}}$ might be the correct critical exponent for the bare lattice version of $\Lam G$.

\subsection*{Remarks about first order transitions}

The discussion above assumes a divergent correlation length at the phase-transition point or line. First order transitions do usually not have such a divergent correlation length, which is one of the reasons one cannot use it to define a continuum field theory, starting from a lattice field theory. However, in the case of quantum gravity the situation can be different, the reason being that the divergent correlation length we have been discussing, coming from the point-point correlation function, the continuum version of which is defined by Eq.~\rf{dis1}, is not necessarily related to the order parameter that defines the first order transition. In such a case, we might observe a double peak in the probability distribution of the order parameter that classifies the transition as first order. We have such a situation in the case of the $A \mi C_{\rm dS}$ transition. The maxima of these peaks for a finite $N_4$ will be located at $\kp_0^{(1)}(N_4) < \kp_0^{(2)}(N_4)$, and they do not merge for $N_4 \to \infty$. Coming from phase $C_{\rm dS}$ we will first meet $\kp_0^{(1)}(N_4)$ and this line will then serve as the pseudo-critical line discussed above, and the correlation length $\xi = N_4^{1/4}$ related to the two-point function can still diverge for $N_4 \to \infty$ in the way discussed above. This is the scenario that we have in mind related to the $A \mi C_{\rm dS}$ transition.

We can of course force the system to cross the pseudo-critical line $\kp_0^{(1)}(N_4)$ and move towards the other pseudo-critical line $\kp_0^{(2)}(N_4)$ that would be met from phase $A$. During such a forced motion the geometries would be mixtures of phase $C_{\rm dS}$ and phase $A$ geometries, and one would not have the phase $C_{\rm dS}$ effective action with the finite-size scaling that we have used and that we have related to the FRG action.

\bibliography{CDT-FRG}% Produces the bibliography via BibTeX.

\begin{thebibliography}{38}%
\makeatletter
\providecommand \@ifxundefined [1]{%
 \@ifx{#1\undefined}
}%
\providecommand \@ifnum [1]{%
 \ifnum #1\expandafter \@firstoftwo
 \else \expandafter \@secondoftwo
 \fi
}%
\providecommand \@ifx [1]{%
 \ifx #1\expandafter \@firstoftwo
 \else \expandafter \@secondoftwo
 \fi
}%
\providecommand \natexlab [1]{#1}%
\providecommand \enquote  [1]{``#1''}%
\providecommand \bibnamefont  [1]{#1}%
\providecommand \bibfnamefont [1]{#1}%
\providecommand \citenamefont [1]{#1}%
\providecommand \href@noop [0]{\@secondoftwo}%
\providecommand \href [0]{\begingroup \@sanitize@url \@href}%
\providecommand \@href[1]{\@@startlink{#1}\@@href}%
\providecommand \@@href[1]{\endgroup#1\@@endlink}%
\providecommand \@sanitize@url [0]{\catcode `\\12\catcode `\$12\catcode `\&12\catcode `\#12\catcode `\^12\catcode `\_12\catcode `\%12\relax}%
\providecommand \@@startlink[1]{}%
\providecommand \@@endlink[0]{}%
\providecommand \url  [0]{\begingroup\@sanitize@url \@url }%
\providecommand \@url [1]{\endgroup\@href {#1}{\urlprefix }}%
\providecommand \urlprefix  [0]{URL }%
\providecommand \Eprint [0]{\href }%
\providecommand \doibase [0]{https://doi.org/}%
\providecommand \selectlanguage [0]{\@gobble}%
\providecommand \bibinfo  [0]{\@secondoftwo}%
\providecommand \bibfield  [0]{\@secondoftwo}%
\providecommand \translation [1]{[#1]}%
\providecommand \BibitemOpen [0]{}%
\providecommand \bibitemStop [0]{}%
\providecommand \bibitemNoStop [0]{.\EOS\space}%
\providecommand \EOS [0]{\spacefactor3000\relax}%
\providecommand \BibitemShut  [1]{\csname bibitem#1\endcsname}%
\let\auto@bib@innerbib\@empty
%</preamble>
\bibitem [{\citenamefont {Ambjorn}\ \emph {et~al.}(2012{\natexlab{a}})\citenamefont {Ambjorn}, \citenamefont {Goerlich}, \citenamefont {Jurkiewicz},\ and\ \citenamefont {Loll}}]{review1}%
  \BibitemOpen
  \bibfield  {author} {\bibinfo {author} {\bibfnamefont {J.}~\bibnamefont {Ambjorn}}, \bibinfo {author} {\bibfnamefont {A.}~\bibnamefont {Goerlich}}, \bibinfo {author} {\bibfnamefont {J.}~\bibnamefont {Jurkiewicz}},\ and\ \bibinfo {author} {\bibfnamefont {R.}~\bibnamefont {Loll}},\ }\bibfield  {title} {\bibinfo {title} {{Nonperturbative Quantum Gravity}},\ }\href {https://doi.org/10.1016/j.physrep.2012.03.007} {\bibfield  {journal} {\bibinfo  {journal} {Phys. Rept.}\ }\textbf {\bibinfo {volume} {519}},\ \bibinfo {pages} {127} (\bibinfo {year} {2012}{\natexlab{a}})},\ \Eprint {https://arxiv.org/abs/1203.3591} {arXiv:1203.3591 [hep-th]} \BibitemShut {NoStop}%
\bibitem [{\citenamefont {Loll}(2020)}]{review2}%
  \BibitemOpen
  \bibfield  {author} {\bibinfo {author} {\bibfnamefont {R.}~\bibnamefont {Loll}},\ }\bibfield  {title} {\bibinfo {title} {{Quantum Gravity from Causal Dynamical Triangulations: A Review}},\ }\href {https://doi.org/10.1088/1361-6382/ab57c7} {\bibfield  {journal} {\bibinfo  {journal} {Class. Quant. Grav.}\ }\textbf {\bibinfo {volume} {37}},\ \bibinfo {pages} {013002} (\bibinfo {year} {2020})},\ \Eprint {https://arxiv.org/abs/1905.08669} {arXiv:1905.08669 [hep-th]} \BibitemShut {NoStop}%
\bibitem [{\citenamefont {Weinberg}(1980)}]{weinberg}%
  \BibitemOpen
  \bibfield  {author} {\bibinfo {author} {\bibfnamefont {S.}~\bibnamefont {Weinberg}},\ }\href@noop {} {\emph {\bibinfo {title} {{General Relativity}: {An Einstein Centenary Survey}}}}\ (\bibinfo {year} {1980})\ pp.\ \bibinfo {pages} {790--831}\BibitemShut {NoStop}%
\bibitem [{\citenamefont {Codello}\ \emph {et~al.}(2009)\citenamefont {Codello}, \citenamefont {Percacci},\ and\ \citenamefont {Rahmede}}]{review3}%
  \BibitemOpen
  \bibfield  {author} {\bibinfo {author} {\bibfnamefont {A.}~\bibnamefont {Codello}}, \bibinfo {author} {\bibfnamefont {R.}~\bibnamefont {Percacci}},\ and\ \bibinfo {author} {\bibfnamefont {C.}~\bibnamefont {Rahmede}},\ }\bibfield  {title} {\bibinfo {title} {Investigating the ultraviolet properties of gravity with a wilsonian renormalization group equation},\ }\href {https://doi.org/https://doi.org/10.1016/j.aop.2008.08.008} {\bibfield  {journal} {\bibinfo  {journal} {Annals of Physics}\ }\textbf {\bibinfo {volume} {324}},\ \bibinfo {pages} {414} (\bibinfo {year} {2009})}\BibitemShut {NoStop}%
\bibitem [{\citenamefont {Reuter}\ and\ \citenamefont {Saueressig}(2012)}]{review4}%
  \BibitemOpen
  \bibfield  {author} {\bibinfo {author} {\bibfnamefont {M.}~\bibnamefont {Reuter}}\ and\ \bibinfo {author} {\bibfnamefont {F.}~\bibnamefont {Saueressig}},\ }\bibfield  {title} {\bibinfo {title} {Quantum einstein gravity},\ }\href {https://doi.org/10.1088/1367-2630/14/5/055022} {\bibfield  {journal} {\bibinfo  {journal} {New Journal of Physics}\ }\textbf {\bibinfo {volume} {14}},\ \bibinfo {pages} {055022} (\bibinfo {year} {2012})}\BibitemShut {NoStop}%
\bibitem [{\citenamefont {Reuter}\ and\ \citenamefont {Saueressig}(2019)}]{review5}%
  \BibitemOpen
  \bibfield  {author} {\bibinfo {author} {\bibfnamefont {M.}~\bibnamefont {Reuter}}\ and\ \bibinfo {author} {\bibfnamefont {F.}~\bibnamefont {Saueressig}},\ }\href@noop {} {\emph {\bibinfo {title} {Quantum Gravity and the Functional Renormalization Group: The Road towards Asymptotic Safety}}},\ Cambridge Monographs on Mathematical Physics\ (\bibinfo  {publisher} {Cambridge University Press},\ \bibinfo {year} {2019})\BibitemShut {NoStop}%
\bibitem [{\citenamefont {Stelle}(1977)}]{stelle}%
  \BibitemOpen
  \bibfield  {author} {\bibinfo {author} {\bibfnamefont {K.~S.}\ \bibnamefont {Stelle}},\ }\bibfield  {title} {\bibinfo {title} {Renormalization of higher-derivative quantum gravity},\ }\href {https://doi.org/10.1103/PhysRevD.16.953} {\bibfield  {journal} {\bibinfo  {journal} {Phys. Rev. D}\ }\textbf {\bibinfo {volume} {16}},\ \bibinfo {pages} {953} (\bibinfo {year} {1977})}\BibitemShut {NoStop}%
\bibitem [{\citenamefont {Ambjorn}\ \emph {et~al.}(2001)\citenamefont {Ambjorn}, \citenamefont {Jurkiewicz},\ and\ \citenamefont {Loll}}]{ajl2001}%
  \BibitemOpen
  \bibfield  {author} {\bibinfo {author} {\bibfnamefont {J.}~\bibnamefont {Ambjorn}}, \bibinfo {author} {\bibfnamefont {J.}~\bibnamefont {Jurkiewicz}},\ and\ \bibinfo {author} {\bibfnamefont {R.}~\bibnamefont {Loll}},\ }\bibfield  {title} {\bibinfo {title} {{Dynamically triangulating Lorentzian quantum gravity}},\ }\href {https://doi.org/10.1016/S0550-3213(01)00297-8} {\bibfield  {journal} {\bibinfo  {journal} {Nucl. Phys. B}\ }\textbf {\bibinfo {volume} {610}},\ \bibinfo {pages} {347} (\bibinfo {year} {2001})},\ \Eprint {https://arxiv.org/abs/hep-th/0105267} {arXiv:hep-th/0105267} \BibitemShut {NoStop}%
\bibitem [{\citenamefont {Ferrero}\ and\ \citenamefont {Reuter}(2021)}]{fr1}%
  \BibitemOpen
  \bibfield  {author} {\bibinfo {author} {\bibfnamefont {R.}~\bibnamefont {Ferrero}}\ and\ \bibinfo {author} {\bibfnamefont {M.}~\bibnamefont {Reuter}},\ }\bibfield  {title} {\bibinfo {title} {Towards a geometrization of renormalization group histories in asymptotic safety},\ }\bibfield  {journal} {\bibinfo  {journal} {Universe}\ }\textbf {\bibinfo {volume} {7}},\ \href {https://doi.org/10.3390/universe7050125} {10.3390/universe7050125} (\bibinfo {year} {2021})\BibitemShut {NoStop}%
\bibitem [{\citenamefont {Ferrero}\ and\ \citenamefont {Reuter}(2022)}]{fr2}%
  \BibitemOpen
  \bibfield  {author} {\bibinfo {author} {\bibfnamefont {R.}~\bibnamefont {Ferrero}}\ and\ \bibinfo {author} {\bibfnamefont {M.}~\bibnamefont {Reuter}},\ }\bibfield  {title} {\bibinfo {title} {The spectral geometry of de sitter space in asymptotic safety},\ }\href {https://doi.org/10.1007/JHEP08(2022)040} {\bibfield  {journal} {\bibinfo  {journal} {J. High Energ. Phys}\ } (\bibinfo {year} {2022})}\BibitemShut {NoStop}%
\bibitem [{\citenamefont {Knorr}\ and\ \citenamefont {Saueressig}(2018)}]{knorr-frank}%
  \BibitemOpen
  \bibfield  {author} {\bibinfo {author} {\bibfnamefont {B.}~\bibnamefont {Knorr}}\ and\ \bibinfo {author} {\bibfnamefont {F.}~\bibnamefont {Saueressig}},\ }\bibfield  {title} {\bibinfo {title} {Towards reconstructing the quantum effective action of gravity},\ }\href {https://doi.org/10.1103/PhysRevLett.121.161304} {\bibfield  {journal} {\bibinfo  {journal} {Phys. Rev. Lett.}\ }\textbf {\bibinfo {volume} {121}},\ \bibinfo {pages} {161304} (\bibinfo {year} {2018})}\BibitemShut {NoStop}%
\bibitem [{\citenamefont {Kawai}\ and\ \citenamefont {Ohta}(2023)}]{kawai}%
  \BibitemOpen
  \bibfield  {author} {\bibinfo {author} {\bibfnamefont {H.}~\bibnamefont {Kawai}}\ and\ \bibinfo {author} {\bibfnamefont {N.}~\bibnamefont {Ohta}},\ }\bibfield  {title} {\bibinfo {title} {{Wave function renormalization and flow of couplings in asymptotically safe quantum gravity}},\ }\href {https://doi.org/10.1103/PhysRevD.107.126025} {\bibfield  {journal} {\bibinfo  {journal} {Phys. Rev. D}\ }\textbf {\bibinfo {volume} {107}},\ \bibinfo {pages} {126025} (\bibinfo {year} {2023})},\ \Eprint {https://arxiv.org/abs/2305.10591} {arXiv:2305.10591 [hep-th]} \BibitemShut {NoStop}%
\bibitem [{\citenamefont {Baldazzi}\ and\ \citenamefont {Falls}(2021)}]{kevin}%
  \BibitemOpen
  \bibfield  {author} {\bibinfo {author} {\bibfnamefont {A.}~\bibnamefont {Baldazzi}}\ and\ \bibinfo {author} {\bibfnamefont {K.}~\bibnamefont {Falls}},\ }\bibfield  {title} {\bibinfo {title} {{Essential Quantum Einstein Gravity}},\ }\href {https://doi.org/10.3390/universe7080294} {\bibfield  {journal} {\bibinfo  {journal} {Universe}\ }\textbf {\bibinfo {volume} {07}},\ \bibinfo {pages} {294} (\bibinfo {year} {2021})},\ \Eprint {https://arxiv.org/abs/2107.00671} {arXiv:2107.00671 [hep-th]} \BibitemShut {NoStop}%
\bibitem [{\citenamefont {Ambjorn}\ \emph {et~al.}(2008)\citenamefont {Ambjorn}, \citenamefont {Gorlich}, \citenamefont {Jurkiewicz},\ and\ \citenamefont {Loll}}]{agjl1}%
  \BibitemOpen
  \bibfield  {author} {\bibinfo {author} {\bibfnamefont {J.}~\bibnamefont {Ambjorn}}, \bibinfo {author} {\bibfnamefont {A.}~\bibnamefont {Gorlich}}, \bibinfo {author} {\bibfnamefont {J.}~\bibnamefont {Jurkiewicz}},\ and\ \bibinfo {author} {\bibfnamefont {R.}~\bibnamefont {Loll}},\ }\bibfield  {title} {\bibinfo {title} {{The Nonperturbative Quantum de Sitter Universe}},\ }\href {https://doi.org/10.1103/PhysRevD.78.063544} {\bibfield  {journal} {\bibinfo  {journal} {Phys. Rev. D}\ }\textbf {\bibinfo {volume} {78}},\ \bibinfo {pages} {063544} (\bibinfo {year} {2008})},\ \Eprint {https://arxiv.org/abs/0807.4481} {arXiv:0807.4481 [hep-th]} \BibitemShut {NoStop}%
\bibitem [{\citenamefont {Montvay}\ and\ \citenamefont {Munster}(1997)}]{munster}%
  \BibitemOpen
  \bibfield  {author} {\bibinfo {author} {\bibfnamefont {I.}~\bibnamefont {Montvay}}\ and\ \bibinfo {author} {\bibfnamefont {G.}~\bibnamefont {Munster}},\ }\href {https://doi.org/10.1017/CBO9780511470783} {\emph {\bibinfo {title} {{Quantum fields on a lattice}}}},\ Cambridge Monographs on Mathematical Physics\ (\bibinfo  {publisher} {Cambridge University Press},\ \bibinfo {year} {1997})\BibitemShut {NoStop}%
\bibitem [{\citenamefont {Ambjorn}\ and\ \citenamefont {Watabiki}(1995)}]{aw}%
  \BibitemOpen
  \bibfield  {author} {\bibinfo {author} {\bibfnamefont {J.}~\bibnamefont {Ambjorn}}\ and\ \bibinfo {author} {\bibfnamefont {Y.}~\bibnamefont {Watabiki}},\ }\bibfield  {title} {\bibinfo {title} {{Scaling in quantum gravity}},\ }\href {https://doi.org/10.1016/0550-3213(95)00154-K} {\bibfield  {journal} {\bibinfo  {journal} {Nucl. Phys. B}\ }\textbf {\bibinfo {volume} {445}},\ \bibinfo {pages} {129} (\bibinfo {year} {1995})},\ \Eprint {https://arxiv.org/abs/hep-th/9501049} {arXiv:hep-th/9501049} \BibitemShut {NoStop}%
\bibitem [{\citenamefont {Ambj\o{}rn}\ \emph {et~al.}(2005)\citenamefont {Ambj\o{}rn}, \citenamefont {Durhuus},\ and\ \citenamefont {Jonsson}}]{book}%
  \BibitemOpen
  \bibfield  {author} {\bibinfo {author} {\bibfnamefont {J.}~\bibnamefont {Ambj\o{}rn}}, \bibinfo {author} {\bibfnamefont {B.}~\bibnamefont {Durhuus}},\ and\ \bibinfo {author} {\bibfnamefont {T.}~\bibnamefont {Jonsson}},\ }\href {https://doi.org/10.1017/CBO9780511524417} {\emph {\bibinfo {title} {{Quantum Geometry}: {A Statistical Field Theory Approach}}}},\ Cambridge Monographs on Mathematical Physics\ (\bibinfo  {publisher} {Cambridge Univ. Press},\ \bibinfo {address} {Cambridge, UK},\ \bibinfo {year} {2005})\BibitemShut {NoStop}%
\bibitem [{\citenamefont {Ambjorn}\ and\ \citenamefont {Loll}(1998)}]{aletal}%
  \BibitemOpen
  \bibfield  {author} {\bibinfo {author} {\bibfnamefont {J.}~\bibnamefont {Ambjorn}}\ and\ \bibinfo {author} {\bibfnamefont {R.}~\bibnamefont {Loll}},\ }\bibfield  {title} {\bibinfo {title} {{Nonperturbative Lorentzian quantum gravity, causality and topology change}},\ }\href {https://doi.org/10.1016/S0550-3213(98)00692-0} {\bibfield  {journal} {\bibinfo  {journal} {Nucl. Phys. B}\ }\textbf {\bibinfo {volume} {536}},\ \bibinfo {pages} {407} (\bibinfo {year} {1998})},\ \Eprint {https://arxiv.org/abs/hep-th/9805108} {arXiv:hep-th/9805108} \BibitemShut {NoStop}%
\bibitem [{\citenamefont {Ambjorn}\ \emph {et~al.}(2005)\citenamefont {Ambjorn}, \citenamefont {Jurkiewicz},\ and\ \citenamefont {Loll}}]{hausdorff4}%
  \BibitemOpen
  \bibfield  {author} {\bibinfo {author} {\bibfnamefont {J.}~\bibnamefont {Ambjorn}}, \bibinfo {author} {\bibfnamefont {J.}~\bibnamefont {Jurkiewicz}},\ and\ \bibinfo {author} {\bibfnamefont {R.}~\bibnamefont {Loll}},\ }\bibfield  {title} {\bibinfo {title} {{Reconstructing the universe}},\ }\href {https://doi.org/10.1103/PhysRevD.72.064014} {\bibfield  {journal} {\bibinfo  {journal} {Phys. Rev. D}\ }\textbf {\bibinfo {volume} {72}},\ \bibinfo {pages} {064014} (\bibinfo {year} {2005})},\ \Eprint {https://arxiv.org/abs/hep-th/0505154} {arXiv:hep-th/0505154} \BibitemShut {NoStop}%
\bibitem [{\citenamefont {Ambjorn}\ \emph {et~al.}(2012{\natexlab{b}})\citenamefont {Ambjorn}, \citenamefont {Jordan}, \citenamefont {Jurkiewicz},\ and\ \citenamefont {Loll}}]{cdt-finitesize}%
  \BibitemOpen
  \bibfield  {author} {\bibinfo {author} {\bibfnamefont {J.}~\bibnamefont {Ambjorn}}, \bibinfo {author} {\bibfnamefont {S.}~\bibnamefont {Jordan}}, \bibinfo {author} {\bibfnamefont {J.}~\bibnamefont {Jurkiewicz}},\ and\ \bibinfo {author} {\bibfnamefont {R.}~\bibnamefont {Loll}},\ }\bibfield  {title} {\bibinfo {title} {{Second- and First-Order Phase Transitions in CDT}},\ }\href {https://doi.org/10.1103/PhysRevD.85.124044} {\bibfield  {journal} {\bibinfo  {journal} {Phys. Rev. D}\ }\textbf {\bibinfo {volume} {85}},\ \bibinfo {pages} {124044} (\bibinfo {year} {2012}{\natexlab{b}})},\ \Eprint {https://arxiv.org/abs/1205.1229} {arXiv:1205.1229 [hep-th]} \BibitemShut {NoStop}%
\bibitem [{\citenamefont {Ambjorn}\ \emph {et~al.}(2020)\citenamefont {Ambjorn}, \citenamefont {Gizbert-Studnicki}, \citenamefont {G\"orlich}, \citenamefont {Jurkiewicz},\ and\ \citenamefont {Loll}}]{reviewfrontier}%
  \BibitemOpen
  \bibfield  {author} {\bibinfo {author} {\bibfnamefont {J.}~\bibnamefont {Ambjorn}}, \bibinfo {author} {\bibfnamefont {J.}~\bibnamefont {Gizbert-Studnicki}}, \bibinfo {author} {\bibfnamefont {A.}~\bibnamefont {G\"orlich}}, \bibinfo {author} {\bibfnamefont {J.}~\bibnamefont {Jurkiewicz}},\ and\ \bibinfo {author} {\bibfnamefont {R.}~\bibnamefont {Loll}},\ }\bibfield  {title} {\bibinfo {title} {{Renormalization in quantum theories of geometry}},\ }\href {https://doi.org/10.3389/fphy.2020.00247} {\bibfield  {journal} {\bibinfo  {journal} {Front. in Phys.}\ }\textbf {\bibinfo {volume} {8}},\ \bibinfo {pages} {247} (\bibinfo {year} {2020})},\ \Eprint {https://arxiv.org/abs/2002.01693} {arXiv:2002.01693 [hep-th]} \BibitemShut {NoStop}%
\bibitem [{\citenamefont {Ambjorn}\ \emph {et~al.}(2014)\citenamefont {Ambjorn}, \citenamefont {G\"orlich}, \citenamefont {Jurkiewicz}, \citenamefont {Kreienbuehl},\ and\ \citenamefont {Loll}}]{cdt-rgf}%
  \BibitemOpen
  \bibfield  {author} {\bibinfo {author} {\bibfnamefont {J.}~\bibnamefont {Ambjorn}}, \bibinfo {author} {\bibfnamefont {A.}~\bibnamefont {G\"orlich}}, \bibinfo {author} {\bibfnamefont {J.}~\bibnamefont {Jurkiewicz}}, \bibinfo {author} {\bibfnamefont {A.}~\bibnamefont {Kreienbuehl}},\ and\ \bibinfo {author} {\bibfnamefont {R.}~\bibnamefont {Loll}},\ }\bibfield  {title} {\bibinfo {title} {{Renormalization Group Flow in CDT}},\ }\href {https://doi.org/10.1088/0264-9381/31/16/165003} {\bibfield  {journal} {\bibinfo  {journal} {Class. Quant. Grav.}\ }\textbf {\bibinfo {volume} {31}},\ \bibinfo {pages} {165003} (\bibinfo {year} {2014})},\ \Eprint {https://arxiv.org/abs/1405.4585} {arXiv:1405.4585 [hep-th]} \BibitemShut {NoStop}%
\bibitem [{\citenamefont {Christiansen}\ \emph {et~al.}(2016)\citenamefont {Christiansen}, \citenamefont {Knorr}, \citenamefont {Pawlowski},\ and\ \citenamefont {Rodigast}}]{pawlowski}%
  \BibitemOpen
  \bibfield  {author} {\bibinfo {author} {\bibfnamefont {N.}~\bibnamefont {Christiansen}}, \bibinfo {author} {\bibfnamefont {B.}~\bibnamefont {Knorr}}, \bibinfo {author} {\bibfnamefont {J.~M.}\ \bibnamefont {Pawlowski}},\ and\ \bibinfo {author} {\bibfnamefont {A.}~\bibnamefont {Rodigast}},\ }\bibfield  {title} {\bibinfo {title} {{Global Flows in Quantum Gravity}},\ }\href {https://doi.org/10.1103/PhysRevD.93.044036} {\bibfield  {journal} {\bibinfo  {journal} {Phys. Rev. D}\ }\textbf {\bibinfo {volume} {93}},\ \bibinfo {pages} {044036} (\bibinfo {year} {2016})},\ \Eprint {https://arxiv.org/abs/1403.1232} {arXiv:1403.1232 [hep-th]} \BibitemShut {NoStop}%
\bibitem [{\citenamefont {Saueressig}\ and\ \citenamefont {Wang}(2023)}]{frank-w}%
  \BibitemOpen
  \bibfield  {author} {\bibinfo {author} {\bibfnamefont {F.}~\bibnamefont {Saueressig}}\ and\ \bibinfo {author} {\bibfnamefont {J.}~\bibnamefont {Wang}},\ }\bibfield  {title} {\bibinfo {title} {{Foliated asymptotically safe gravity in the fluctuation approach}},\ }\href {https://doi.org/10.1007/JHEP09(2023)064} {\bibfield  {journal} {\bibinfo  {journal} {JHEP}\ }\textbf {\bibinfo {volume} {09}},\ \bibinfo {pages} {064}},\ \Eprint {https://arxiv.org/abs/2306.10408} {arXiv:2306.10408 [hep-th]} \BibitemShut {NoStop}%
\bibitem [{\citenamefont {Ambj\o{}rn}\ \emph {et~al.}(2019)\citenamefont {Ambj\o{}rn}, \citenamefont {Coumbe}, \citenamefont {Gizbert-Studnicki}, \citenamefont {G\"orlich},\ and\ \citenamefont {Jurkiewicz}}]{Ambjorn:2019pkp}%
  \BibitemOpen
  \bibfield  {author} {\bibinfo {author} {\bibfnamefont {J.}~\bibnamefont {Ambj\o{}rn}}, \bibinfo {author} {\bibfnamefont {D.}~\bibnamefont {Coumbe}}, \bibinfo {author} {\bibfnamefont {J.}~\bibnamefont {Gizbert-Studnicki}}, \bibinfo {author} {\bibfnamefont {A.}~\bibnamefont {G\"orlich}},\ and\ \bibinfo {author} {\bibfnamefont {J.}~\bibnamefont {Jurkiewicz}},\ }\bibfield  {title} {\bibinfo {title} {{Critical Phenomena in Causal Dynamical Triangulations}},\ }\href {https://doi.org/10.1088/1361-6382/ab4184} {\bibfield  {journal} {\bibinfo  {journal} {Class. Quant. Grav.}\ }\textbf {\bibinfo {volume} {36}},\ \bibinfo {pages} {224001} (\bibinfo {year} {2019})},\ \Eprint {https://arxiv.org/abs/1904.05755} {arXiv:1904.05755 [hep-th]} \BibitemShut {NoStop}%
\bibitem [{\citenamefont {Ho\ifmmode~\check{r}\else \v{r}\fi{}ava}(2009)}]{horava}%
  \BibitemOpen
  \bibfield  {author} {\bibinfo {author} {\bibfnamefont {P.}~\bibnamefont {Ho\ifmmode~\check{r}\else \v{r}\fi{}ava}},\ }\bibfield  {title} {\bibinfo {title} {Quantum gravity at a lifshitz point},\ }\href {https://doi.org/10.1103/PhysRevD.79.084008} {\bibfield  {journal} {\bibinfo  {journal} {Phys. Rev. D}\ }\textbf {\bibinfo {volume} {79}},\ \bibinfo {pages} {084008} (\bibinfo {year} {2009})}\BibitemShut {NoStop}%
\bibitem [{\citenamefont {Coumbe}\ \emph {et~al.}(2016)\citenamefont {Coumbe}, \citenamefont {Gizbert-Studnicki},\ and\ \citenamefont {Jurkiewicz}}]{Coumbe2016}%
  \BibitemOpen
  \bibfield  {author} {\bibinfo {author} {\bibfnamefont {D.~N.}\ \bibnamefont {Coumbe}}, \bibinfo {author} {\bibfnamefont {J.}~\bibnamefont {Gizbert-Studnicki}},\ and\ \bibinfo {author} {\bibfnamefont {J.}~\bibnamefont {Jurkiewicz}},\ }\bibfield  {title} {\bibinfo {title} {{Exploring the new phase transition of CDT}},\ }\href {https://doi.org/10.1007/JHEP02(2016)144} {\bibfield  {journal} {\bibinfo  {journal} {JHEP}\ }\textbf {\bibinfo {volume} {02}},\ \bibinfo {pages} {144}},\ \Eprint {https://arxiv.org/abs/1510.08672} {arXiv:1510.08672 [hep-th]} \BibitemShut {NoStop}%
\bibitem [{\citenamefont {Ambjorn}\ \emph {et~al.}(2017)\citenamefont {Ambjorn}, \citenamefont {Coumbe}, \citenamefont {Gizbert-Studnicki}, \citenamefont {G\"orlich},\ and\ \citenamefont {Jurkiewicz}}]{Coumbe2017}%
  \BibitemOpen
  \bibfield  {author} {\bibinfo {author} {\bibfnamefont {J.}~\bibnamefont {Ambjorn}}, \bibinfo {author} {\bibfnamefont {D.}~\bibnamefont {Coumbe}}, \bibinfo {author} {\bibfnamefont {J.}~\bibnamefont {Gizbert-Studnicki}}, \bibinfo {author} {\bibfnamefont {A.}~\bibnamefont {G\"orlich}},\ and\ \bibinfo {author} {\bibfnamefont {J.}~\bibnamefont {Jurkiewicz}},\ }\bibfield  {title} {\bibinfo {title} {New higher-order transition in causal dynamical triangulations},\ }\href {https://doi.org/10.1103/PhysRevD.95.124029} {\bibfield  {journal} {\bibinfo  {journal} {Phys. Rev. D}\ }\textbf {\bibinfo {volume} {95}},\ \bibinfo {pages} {124029} (\bibinfo {year} {2017})}\BibitemShut {NoStop}%
\bibitem [{\citenamefont {Falls}\ and\ \citenamefont {Ferrero}( )}]{kevin-renata}%
  \BibitemOpen
  \bibfield  {author} {\bibinfo {author} {\bibfnamefont {K.}~\bibnamefont {Falls}}\ and\ \bibinfo {author} {\bibfnamefont {R.}~\bibnamefont {Ferrero}},\ }\bibfield  {title} {\bibinfo {title} {{To appear}},\ }\href {https://doi.org/~} {\bibfield  {journal} {\bibinfo  {journal} {~}\ }\textbf {\bibinfo {volume} {~}},\ \bibinfo {pages} {~} (\bibinfo {year} {~})},\ \Eprint {https://arxiv.org/abs/~} {~:~ [~]} \BibitemShut {NoStop}%
\bibitem [{\citenamefont {Ambjørn}\ \emph {et~al.}(2011)\citenamefont {Ambjørn}, \citenamefont {Görlich}, \citenamefont {Jurkiewicz}, \citenamefont {Loll}, \citenamefont {Gizbert-Studnicki},\ and\ \citenamefont {Trześniewski}}]{AMBJORN2011144}%
  \BibitemOpen
  \bibfield  {author} {\bibinfo {author} {\bibfnamefont {J.}~\bibnamefont {Ambjørn}}, \bibinfo {author} {\bibfnamefont {A.}~\bibnamefont {Görlich}}, \bibinfo {author} {\bibfnamefont {J.}~\bibnamefont {Jurkiewicz}}, \bibinfo {author} {\bibfnamefont {R.}~\bibnamefont {Loll}}, \bibinfo {author} {\bibfnamefont {J.}~\bibnamefont {Gizbert-Studnicki}},\ and\ \bibinfo {author} {\bibfnamefont {T.}~\bibnamefont {Trześniewski}},\ }\bibfield  {title} {\bibinfo {title} {The semiclassical limit of causal dynamical triangulations},\ }\href {https://doi.org/https://doi.org/10.1016/j.nuclphysb.2011.03.019} {\bibfield  {journal} {\bibinfo  {journal} {Nuclear Physics B}\ }\textbf {\bibinfo {volume} {849}},\ \bibinfo {pages} {144} (\bibinfo {year} {2011})}\BibitemShut {NoStop}%
\bibitem [{\citenamefont {Gizbert-Studnicki}(2023)}]{HandbookJGS}%
  \BibitemOpen
  \bibfield  {author} {\bibinfo {author} {\bibfnamefont {J.}~\bibnamefont {Gizbert-Studnicki}},\ }\bibinfo {title} {Semiclassical and continuum limits of four-dimensional cdt},\ in\ \href {https://doi.org/10.1007/978-981-19-3079-9_95-1} {\emph {\bibinfo {booktitle} {Handbook of Quantum Gravity}}},\ \bibinfo {editor} {edited by\ \bibinfo {editor} {\bibfnamefont {C.}~\bibnamefont {Bambi}}, \bibinfo {editor} {\bibfnamefont {L.}~\bibnamefont {Modesto}},\ and\ \bibinfo {editor} {\bibfnamefont {I.}~\bibnamefont {Shapiro}}}\ (\bibinfo  {publisher} {Springer Nature Singapore},\ \bibinfo {address} {Singapore},\ \bibinfo {year} {2023})\ pp.\ \bibinfo {pages} {1--43}\BibitemShut {NoStop}%
\bibitem [{\citenamefont {Ambjorn}\ \emph {et~al.}(2012{\natexlab{c}})\citenamefont {Ambjorn}, \citenamefont {Goerlich}, \citenamefont {Jurkiewicz},\ and\ \citenamefont {Zhang}}]{Ambjorn:2012kda}%
  \BibitemOpen
  \bibfield  {author} {\bibinfo {author} {\bibfnamefont {J.}~\bibnamefont {Ambjorn}}, \bibinfo {author} {\bibfnamefont {A.~T.}\ \bibnamefont {Goerlich}}, \bibinfo {author} {\bibfnamefont {J.}~\bibnamefont {Jurkiewicz}},\ and\ \bibinfo {author} {\bibfnamefont {H.~G.}\ \bibnamefont {Zhang}},\ }\bibfield  {title} {\bibinfo {title} {{Pseudo-topological transitions in 2D gravity models coupled to massless scalar fields}},\ }\href {https://doi.org/10.1016/j.nuclphysb.2012.05.024} {\bibfield  {journal} {\bibinfo  {journal} {Nucl. Phys. B}\ }\textbf {\bibinfo {volume} {863}},\ \bibinfo {pages} {421} (\bibinfo {year} {2012}{\natexlab{c}})},\ \Eprint {https://arxiv.org/abs/1201.1590} {arXiv:1201.1590 [gr-qc]} \BibitemShut {NoStop}%
\bibitem [{\citenamefont {Ambj\o{}rn}\ \emph {et~al.}(2021)\citenamefont {Ambj\o{}rn}, \citenamefont {Drogosz}, \citenamefont {Gizbert-Studnicki}, \citenamefont {G\"orlich}, \citenamefont {Jurkiewicz},\ and\ \citenamefont {N\'emeth}}]{Ambjorn:2021fkp}%
  \BibitemOpen
  \bibfield  {author} {\bibinfo {author} {\bibfnamefont {J.}~\bibnamefont {Ambj\o{}rn}}, \bibinfo {author} {\bibfnamefont {Z.}~\bibnamefont {Drogosz}}, \bibinfo {author} {\bibfnamefont {J.}~\bibnamefont {Gizbert-Studnicki}}, \bibinfo {author} {\bibfnamefont {A.}~\bibnamefont {G\"orlich}}, \bibinfo {author} {\bibfnamefont {J.}~\bibnamefont {Jurkiewicz}},\ and\ \bibinfo {author} {\bibfnamefont {D.}~\bibnamefont {N\'emeth}},\ }\bibfield  {title} {\bibinfo {title} {{Matter-Driven Change of Spacetime Topology}},\ }\href {https://doi.org/10.1103/PhysRevLett.127.161301} {\bibfield  {journal} {\bibinfo  {journal} {Phys. Rev. Lett.}\ }\textbf {\bibinfo {volume} {127}},\ \bibinfo {pages} {161301} (\bibinfo {year} {2021})},\ \Eprint {https://arxiv.org/abs/2103.00198} {arXiv:2103.00198 [hep-th]} \BibitemShut {NoStop}%
\bibitem [{\citenamefont {Ambjorn}\ and\ \citenamefont {Jurkiewicz}(1992)}]{aj1}%
  \BibitemOpen
  \bibfield  {author} {\bibinfo {author} {\bibfnamefont {J.}~\bibnamefont {Ambjorn}}\ and\ \bibinfo {author} {\bibfnamefont {J.}~\bibnamefont {Jurkiewicz}},\ }\bibfield  {title} {\bibinfo {title} {{Four-dimensional simplicial quantum gravity}},\ }\href {https://doi.org/10.1016/0370-2693(92)90709-D} {\bibfield  {journal} {\bibinfo  {journal} {Phys. Lett. B}\ }\textbf {\bibinfo {volume} {278}},\ \bibinfo {pages} {42} (\bibinfo {year} {1992})}\BibitemShut {NoStop}%
\bibitem [{\citenamefont {Ambjorn}\ and\ \citenamefont {Jurkiewicz}(1995)}]{aj2}%
  \BibitemOpen
  \bibfield  {author} {\bibinfo {author} {\bibfnamefont {J.}~\bibnamefont {Ambjorn}}\ and\ \bibinfo {author} {\bibfnamefont {J.}~\bibnamefont {Jurkiewicz}},\ }\bibfield  {title} {\bibinfo {title} {{Scaling in four-dimensional quantum gravity}},\ }\href {https://doi.org/10.1016/0550-3213(95)00303-A} {\bibfield  {journal} {\bibinfo  {journal} {Nucl. Phys. B}\ }\textbf {\bibinfo {volume} {451}},\ \bibinfo {pages} {643} (\bibinfo {year} {1995})},\ \Eprint {https://arxiv.org/abs/hep-th/9503006} {arXiv:hep-th/9503006} \BibitemShut {NoStop}%
\bibitem [{\citenamefont {Bassler}\ \emph {et~al.}(2021)\citenamefont {Bassler}, \citenamefont {Laiho}, \citenamefont {Schiffer},\ and\ \citenamefont {Unmuth-Yockey}}]{jack}%
  \BibitemOpen
  \bibfield  {author} {\bibinfo {author} {\bibfnamefont {S.}~\bibnamefont {Bassler}}, \bibinfo {author} {\bibfnamefont {J.}~\bibnamefont {Laiho}}, \bibinfo {author} {\bibfnamefont {M.}~\bibnamefont {Schiffer}},\ and\ \bibinfo {author} {\bibfnamefont {J.}~\bibnamefont {Unmuth-Yockey}},\ }\bibfield  {title} {\bibinfo {title} {{The de Sitter Instanton from Euclidean Dynamical Triangulations}},\ }\href {https://doi.org/10.1103/PhysRevD.103.114504} {\bibfield  {journal} {\bibinfo  {journal} {Phys. Rev. D}\ }\textbf {\bibinfo {volume} {103}},\ \bibinfo {pages} {114504} (\bibinfo {year} {2021})},\ \Eprint {https://arxiv.org/abs/2103.06973} {arXiv:2103.06973 [hep-lat]} \BibitemShut {NoStop}%
\bibitem [{\citenamefont {Laiho}\ \emph {et~al.}(2017)\citenamefont {Laiho}, \citenamefont {Bassler}, \citenamefont {Coumbe}, \citenamefont {Du},\ and\ \citenamefont {Neelakanta}}]{jack1}%
  \BibitemOpen
  \bibfield  {author} {\bibinfo {author} {\bibfnamefont {J.}~\bibnamefont {Laiho}}, \bibinfo {author} {\bibfnamefont {S.}~\bibnamefont {Bassler}}, \bibinfo {author} {\bibfnamefont {D.}~\bibnamefont {Coumbe}}, \bibinfo {author} {\bibfnamefont {D.}~\bibnamefont {Du}},\ and\ \bibinfo {author} {\bibfnamefont {J.~T.}\ \bibnamefont {Neelakanta}},\ }\bibfield  {title} {\bibinfo {title} {{Lattice Quantum Gravity and Asymptotic Safety}},\ }\href {https://doi.org/10.1103/PhysRevD.96.064015} {\bibfield  {journal} {\bibinfo  {journal} {Phys. Rev. D}\ }\textbf {\bibinfo {volume} {96}},\ \bibinfo {pages} {064015} (\bibinfo {year} {2017})},\ \Eprint {https://arxiv.org/abs/1604.02745} {arXiv:1604.02745 [hep-th]} \BibitemShut {NoStop}%
\bibitem [{\citenamefont {Luscher}\ and\ \citenamefont {Weisz}(1987)}]{luescher}%
  \BibitemOpen
  \bibfield  {author} {\bibinfo {author} {\bibfnamefont {M.}~\bibnamefont {Luscher}}\ and\ \bibinfo {author} {\bibfnamefont {P.}~\bibnamefont {Weisz}},\ }\bibfield  {title} {\bibinfo {title} {{Scaling Laws and Triviality Bounds in the Lattice phi**4 Theory. 1. One Component Model in the Symmetric Phase}},\ }\href {https://doi.org/10.1016/0550-3213(87)90177-5} {\bibfield  {journal} {\bibinfo  {journal} {Nucl. Phys. B}\ }\textbf {\bibinfo {volume} {290}},\ \bibinfo {pages} {25} (\bibinfo {year} {1987})}\BibitemShut {NoStop}%
\end{thebibliography}%

\end{document}